\documentclass{aa}  

\usepackage{graphicx}
\usepackage{orcidlink}
\usepackage{txfonts}

\usepackage{array}
\newcolumntype{h}{>{\setbox0=\hbox\bgroup}c<{\egroup}@{}}

\begin{document} 

\newcommand{\SiOa}{$^{29}$SiO}
\newcommand{\SiOb}{$^{30}$SiO}
\newcommand{\jone}{{$J$=1$\rightarrow$0}}
\newcommand{\jtwo}{{$J$=2$\rightarrow$1}}
\newcommand{\vone}{$v$=1}
\newcommand{\vtwo}{$v$=2}
\newcommand{\vthr}{$v$=3}
\newcommand{\ls}[1]{\textcolor{orange}{\bfseries #1}}

   \title{\SiOa\ and \SiOb\ \jone\ maser signatures in Galactic AGB stars}

   \subtitle{The impact of third dredge up and turbulence velocity}

   \author{M.O. Lewis
          \inst{1}
          \orcidlink{0000-0002-8069-8060}
          \and
          L.O. Sjouwerman 
          \inst{2}\fnmsep\thanks{LOS is also an adjunct professor at the Department of Physics and Astronomy, University of New Mexico}
          \orcidlink{0000-0003-3096-3062}
          \and
          Y.M. Pihlström 
          \inst{2} \fnmsep\thanks{YMP is also an adjunct professor at the Department of Physics and Astronomy, University of New Mexico}
          \orcidlink{0000-0003-0615-1785}
          \and
          H. J. van Langevelde
          \inst{3,1}
          \orcidlink{0000-0002-0230-5946}
          \and
          R. Bhattacharya
          \inst{2,4} 
          \orcidlink{0009-0007-8229-3036}
          \and
          M.C. Stroh
          \inst{2}
          \orcidlink{0000-0002-3019-4577}   
          }

   \institute{
        Leiden Observatory, Leiden University, PO Box 9513, 2300 RA Leiden, The Netherlands
     \and
        National Radio Astronomy Observatory, Pete V. Domenici Science Operations Center, Socorro, NM 87801, USA
     \and 
        Joint Institute for VLBI ERIC (JIVE), Oude Hoogeveensedijk 4, 7991 PD Dwingeloo, The Netherlands
     \and
        Department of Physics and Astronomy, University of New Mexico, Albuquerque, NM 87131, USA
         }
   \date{Received Feb 20 2026; accepted May 4 2026}

 
  \abstract
   {\jone\ SiO masers at 43 GHz have a well-established distinctive signature in asymptotic giant branch (AGB) stars. $^{28}$SiO transitions typically dominate these spectra with the \textit{v}=1 and \textit{v}=2 emission being especially prominent and ubiquitous. However, with the availability of over 10,000 \jone\ SiO maser spectra from the BAaDE survey, extreme deviations from this typical signature have been identified. Several predictions about enhanced $^{29}$SiO abundances in exotic stars prompt us to catalog the cases where $^{29}$SiO maser emission is extremely enhanced compared to $^{28}$SiO.}
   {Our purpose is to catalog the known cases of 43 GHz spectra dominated by emission from isotopologue transitions (isotopologue-dominated spectra), to explore the commonalities in these sources, and to attempt to explain the cause of such maser signatures. }
   {Our catalog is drawn from SiO maser line ratios in the infrared-color-selected BAaDE survey and supplemented with a literature detection. We extract variability properties and infrared photometry for these sources from the literature.}
   {The BAaDE catalog has cemented the typical signature of 43 GHz SiO masers, showing it is dominated by the \textit{v}=1 and \textit{v}=2 lines, as expected. Thirty-five isotopologue-dominated spectra are identified, meaning that this signature is seen in about 0.2\% of our SiO maser-bearing stars. Their infrared colors are blue compared to other sources of the same period, similar to all sources displaying isotopologue masers. It is clear that the isotopologue-dominated nature of sources is variable, but unclear whether this is tied to stellar phase. 
   
   }
   {A large abundance abnormality among the isotopologue-dominated sources is disfavored as the isotopologue-dominated sources do not appear significantly different from other stars which host isotopologue masers. Maser pumping, affecting the population inversions of specific transitions, can instead explain the enhanced signatures. We posit that isotopologue-dominated spectra appear in AGB stars which have undergone third-dredge up (enhancing the \SiOa\  and \SiOb\ abundance slightly) and which, in addition, are experiencing very low turbulence velocity ($\lesssim$1 km s$^{-1}$), creating a line overlap which pumps the maser transitions very efficiently. }


   \maketitle
%

\section{Introduction}

SiO masers form commonly in the circumstellar envelopes (CSEs) surrounding asymptotic giant branch (AGB) stars. The BAaDE survey obtained over 10,000 SiO maser spectra at 43 GHz \citep{baade}. Such a large sample probes the statistical properties of masers, joining a host of previous observations (\citealt{deguchi08,nakashimaanddeguchi07,rizzo21, baek25} and many more), as these masers have proven to be efficient tracers of line-of-sight velocity throughout the Galaxy.

The origin of the population inversion leading to SiO maser emission has been explained by radiative and collisional pumping models. Radiative pumping models are supported by the observation that the maser flux is correlated with the IR emission \citep{bujarrabal87}, including the variability of the IR emission with a possible 0.2 lag in stellar phase \citep{alcolea99,pardo04, gonidakis13}. Alternatively, on grounds of the roughly spatial co-location of the \textit{v}=1 and \textit{v}=2 emission \citep{miyoshi94, desmurs00, soriaruiz04}, collisional pumping occurring in shocks induced by the expanding and contracting CSE was proposed by \cite{miyoshi94}. However, small differences in the positions of these masers have been used as evidence of radiative pumping \citep{soriaruiz04}. This maser pumping debate is still ongoing, and likely both physical processes play a role.

As maser emission is exponentially amplified, maser line ratios are difficult to relate to physical and chemical characteristics, especially given the variability of this emission, although some relation to molecular abundances is expected. For example, the most basic chemical distinction between AGB stars is whether their CSEs are dominated by carbon (C-rich) or oxygen (O-rich) molecules, and maser species are strongly affected by this chemical division. O-rich stars generally host OH, H$_2$O, and/or SiO masers \citep{reid&moran81}; C-rich stars are more likely to host HCN masers \citep{menten18}. S-type AGB stars, which have \textit{s}-process elements within their circumstellar envelopes, and are generally an intermediate chemical type between O- and C-rich, have shown their own unique maser signatures, for example displaying \textit{v}=2 \jtwo\ SiO masers \citep{bujarrabal96, michael19}.
 
$^{28}$SiO (hereafter just SiO) is the most common isotopologue of SiO, and the maser emission from this molecule is generally the brightest signature in a maser spectrum.

\SiOa\ and \SiOb\ are also stable isotopologues which may display detectable (maser) emission.
The abundance of these three SiO isotopologues was investigated by \cite{monson17} using thermal SiO emission mostly in star-forming regions. They conclude that the isotopologue abundances are roughly stable across the Milky Way, with the $^{28}$SiO/\SiOa\ being $\sim$18 and the $^{28}$SiO/\SiOb\ being $\sim$25 \citep{monson17}. \cite{peng13} study similar ratios in evolved stars, but only for \SiOa\ and \SiOb\ where the emission is unsaturated (and non-maser), also finding that the $^{29}$Si/$^{30}$Si ratio is between 1-1.5. Metallicity and mass also affect the ratios of $^{29}$Si/$^{28}$Si and $^{30}$Si/$^{28}$Si in evolved stars, with higher masses ($\sim$5 M$_{\odot}$) and higher metallicity (Z = 0.02) generally yielding more relative $^{29}$Si and $^{30}$Si \citep{zinner06} in the CSE.

The changes in the relative abundances of silicon isotopes are likely produced by third dredge up (TDU), which brings nucleosynthesis products from the stellar interior into the CSE via large convection cells \citep{herwig05}. Slow neutron capture is taking place in the stellar interior \citep{busso99}; this can produce both $^{29}$Si and $^{30}$Si which subsequently can be brought into the CSE via TDU. 

Abundances of $^{29}$Si and $^{30}$Si are higher than solar values in presolar mainstream SiC grains, which are produced in C-rich AGB stars \citep{lugaro99}, showing that dredge-up events can increase the amount of available $^{29}$Si and $^{30}$Si in an AGB envelope. Specifically, \cite{zinner06} showed that the $^{29}$Si/$^{28}$Si ratio can be increased even before TDU converts the star to C-rich. Near solar metallicity (z=0.01-0.02 but not lower) and higher masses ($\gtrsim$ 5 M$_{\odot}$) are required for the ratio to change before conversion to C-rich chemistry \citep{zinner06}, but these physical parameters are reasonable for stars towards the bulge of the Milky Way, like in our sample. Similarly, \cite{karakas14} investigate TDU in AGB stars of various metallicities, and also find that the solar- and high-metallicity sources at 5 M$_{\odot}$ will experience TDU but remain slightly O-rich (C/O $\sim$0.8 depending on metallicity and helium abundance). Again these higher-metallicity results are relevant for the bulge stars in our sample.

Exotic stars have been postulated to be tied to high \SiOa\, abundances. For example, \cite{para95} suggest looking for bright \SiOa\  \textit{v}=0 maser emission to identify Thorne-\.Zytkow objects, which are red giant stars with neutron star cores. 
However, bright \SiOa\, maser transitions can also be explained without invoking abnormal abundances. 
Quenching and boosting mechanisms driven by line overlaps can also have huge effects on maser flux density regardless of abundance. For example, the \jtwo\ \textit{v}=2 transition of SiO has a relatively low detection rate even in O-rich sources displaying many other SiO maser transitions, while S-type stars more commonly show maser emission from this transition. This is caused by the line overlap with water \citep{olofsson81}. 
O-rich envelopes host enough water for this overlap to strongly quench the \textit{v}=2 maser transition while S-type stars do not. 
Although this example is a case where an abundance change causes a large change in a specific maser signature, given the complicated interplay of many potential overlaps, line broadening caused by e.g., temperature and turbulence velocity can also cause dramatic changes in the maser spectra \citep{cernicharo93, gac97, herpinandbaudry00}.
Overlaps between transitions of SiO, \SiOa, and/or \SiOb\ can cause similar effects especially influencing isotopologue transitions as shown in the models of \cite{gac97} and \cite{cernicharo93}.

Here we report on a rare class of 43 GHz SiO maser spectra which show unusually bright \SiOa\ and/or \SiOb\ maser lines or ``isotopologue-dominated" (iso-dom) spectra. In Section 2 we introduce the main dataset and establish the typical line ratios from which our main catalog will deviate. In Section 3 we give the properties of the iso-dom spectra and establish spectral subsample types among iso-dom sources. In Section 4 we examine other common characteristics (in terms of variability and IR colors) in the sample.
We discuss the likelihood that these sources have undergone TDU, as well as stellar phase trends, and connections to maser models in Section 5. Section 6 gives our conclusions. 

\section{The standard SiO signature at 43 GHz (\jone)}

With $\sim$28000 sources surveyed with the VLA and ALMA, the BAaDE survey is the largest and most homogeneous SiO maser database to date. The database has 18162 spectra of the \jone\ (43 GHz) transitions from VLA data, and 9701 are detected in one or more transitions \citep{baade, lewis21}. The VLA BAaDE data covers six SiO maser lines plus the SiO \textit{v}=0 line, which may show thermal or maser emission, and occasionally is a mix of
both \citep{dike21} (see Table \ref{tab:lines}). The typical observing time was 40 s per source, leading to RMS noise levels $\sim$ 15 mJy per channel. 

 \begin{table}
 \caption{Rest frequencies}\label{tab:lines}
\centering
 \begin{tabular}{c|c}
      Rest frequency & Transition\\
       (MHz) &  (\jone) \\
\noalign{\hrule}\noalign{\smallskip}
      42373.341 & $^{30}$SiO $v$=0 $^{\phantom{p}}$\\
      42519.375 & $^{\phantom{30}}$SiO $v$=3 $^{\phantom{p}}$\\
      42583.827 & $^{29}$SiO $v$=1 $^{\phantom{p}}$\\
      42820.570 & $^{\phantom{30}}$SiO $v$=2 $^{p}$\\
      42879.941 & $^{29}$SiO $v$=0 $^{*}$\\
      43122.090 & $^{\phantom{30}}$SiO $v$=1 $^{p}$\\
      43423.853 & $^{\phantom{30}}$SiO $v$=0 $^{t}$\\
 \end{tabular}
 \tablefoot{Rest frequencies  of the seven transitions that are observed by the VLA portion of the BAaDE survey. The primary lines are marked (p), as well as the isotopologue line (*) used to determine whether a star is isotopologue-dominated. The SiO \textit{v}=0 line, marked (t), is not always a maser and can be sometimes thermal or a mix of both.}
  \end{table}

For all sources all transitions are observed simultaneously.
Fifty-one BAaDE sources have been observed for SiO \jone\ in a second epoch \citep{lewis20b}; all other data is single epoch.
All surveyed sources were drawn from the specific iiia color region of an MSX color-color diagram \citep{scc09} in order to optimize detection rate; thus both very blue (with minimal envelopes) and very red (OH/IR-like) objects were not included (see \cite{scc09}, \cite{lewis20a}, and Bhattacharya et al.\,(in prep.) for more details on color selection and its effect on the sample).

With these homogeneous, simultaneous multi-maser-line data the standard SiO signature at 43 GHz is clearly established: i) the SiO \textit{v}=1 and \textit{v}=2 masers are almost always the brightest in a spectrum, ii) they are almost always detected as a pair, iii) the line ratio between the two is nearly unity ($R_{\rm{v=2/v=1}}$ = 0.93$\pm$0.22), and iv) detections of other masers almost always occur in spectra with these two lines present \citep{lewis24}. We refer to the SiO \textit{v}=1 and \textit{v}=2 as the primary lines henceforth. The primary lines are on average 6 times brighter than the next brightest maser line (which is typically the SiO \textit{v}=3 transition). 

Focusing on the isotopologue transitions, the \SiOa\ \textit{v}=0 line is on average 9 times fainter than the primary lines. It is detected in $\sim$ 8 \% of BAaDE sources. The \SiOb\ \textit{v}=0 and \SiOa\ \textit{v}=1 lines are also probed but are detected in only 223 and 19 sources, respectively. See \cite{baade} for data related to these signatures and \cite{lewis24} for basic detection and line ratio statistics.
This picture is supported by a host of other observations at \jone, some targeting rare lines \citep{barcia89, cho96b, rizzo21, baek25} and many others only targeting the SiO \textit{v}=1 and/or \textit{v}=2 \citep{deguchi00, deguchi04, fujii06, kim14, gomezgarrido20}.

Low detection rates of the isotopologue transitions can partially be explained by sensitivity, given the differences in the typical flux densities between transitions.
Given that the \SiOa\ \textit{v}=0 is on average the brightest isotopologue transition, we could assume that isotopologue lines are only likely detectable in sources with primary line detections 9 times over our normal detection threshold of 4.7 times the RMS. This constraint thus depends on the RMS per source (Table \ref{tab:baadeoverview}). Typically a primary line detection in the range 600 to 650 mJy/channel is needed to expect a detection of a potential isotopologue maser.
Indeed the \SiOa\ \textit{v}=0 is found in about 37\% of sources (1393 sources) with SiO \textit{v}=1 emission over 9 times above the detection threshold.
Thus it is clear that emission from this transition is quite common and that there are sensitivity effects limiting the number of detections of isotopologue lines.

\section{Isotopologue-dominated 43 GHz spectra}
As stated, isotopologue maser emission is quite ubiquitous, but here we draw attention to a rare occurrence: where the maser emission from an isotopologue is brighter than the SiO \textit{v}=1 in the 43 GHz spectrum. Herein these are referred to as iso-dom maser spectra. 
Such extreme outlier ratios can be found especially involving the primary lines and the \SiOa\ \textit{v}=0.

We present 36 sources which show anomalously high \SiOa\ maser emission in at least one epoch in Table \ref{tab:otherobs}. Thirty-five are reported from the BAaDE data, and we include an additional source, S Cas, from \cite{rizzo21} and \cite {son25} which shows the same maser signature. All sources where the \SiOa\ \textit{v}=0 emission is brighter than the SiO \textit{v}=1 or the \textit{v}=2 are included in our catalog, and we include one additional source with bright \SiOa\ \textit{v}=1 emission but no \SiOa\ \textit{v}=0 emission. In Table \ref{tab:otherobs} we present the source names across various IR and variability catalogs collected from 2MASS, MSX, WISE, AKARI, Gaia, and OGLE. The spectra of the BAaDE sources are presented in the Appendix, labeled by their BAaDE names.

Periods are determined for many of these sources by  OGLE \citep{udalski15, iwanek22}, Gaia \citep{gaia16, gaia23}, and \cite{urago20}, confirming their AGB and Mira variable star status. We extract the phase of the maser observation using OGLE photometry when available, or Gaia photometry 
in the absence of an alternative.
OGLE data is preferred because of the long time baseline, and Gaia is only used 
if multi-epoch data is present.
In Table 2 we list the periods, phases, and our categorization of the maser spectrum. In Table 3 we list the BAaDE flux densities of the detected maser lines in each source.

\begin{figure*}
    \centering
    \includegraphics[width=0.95\linewidth]{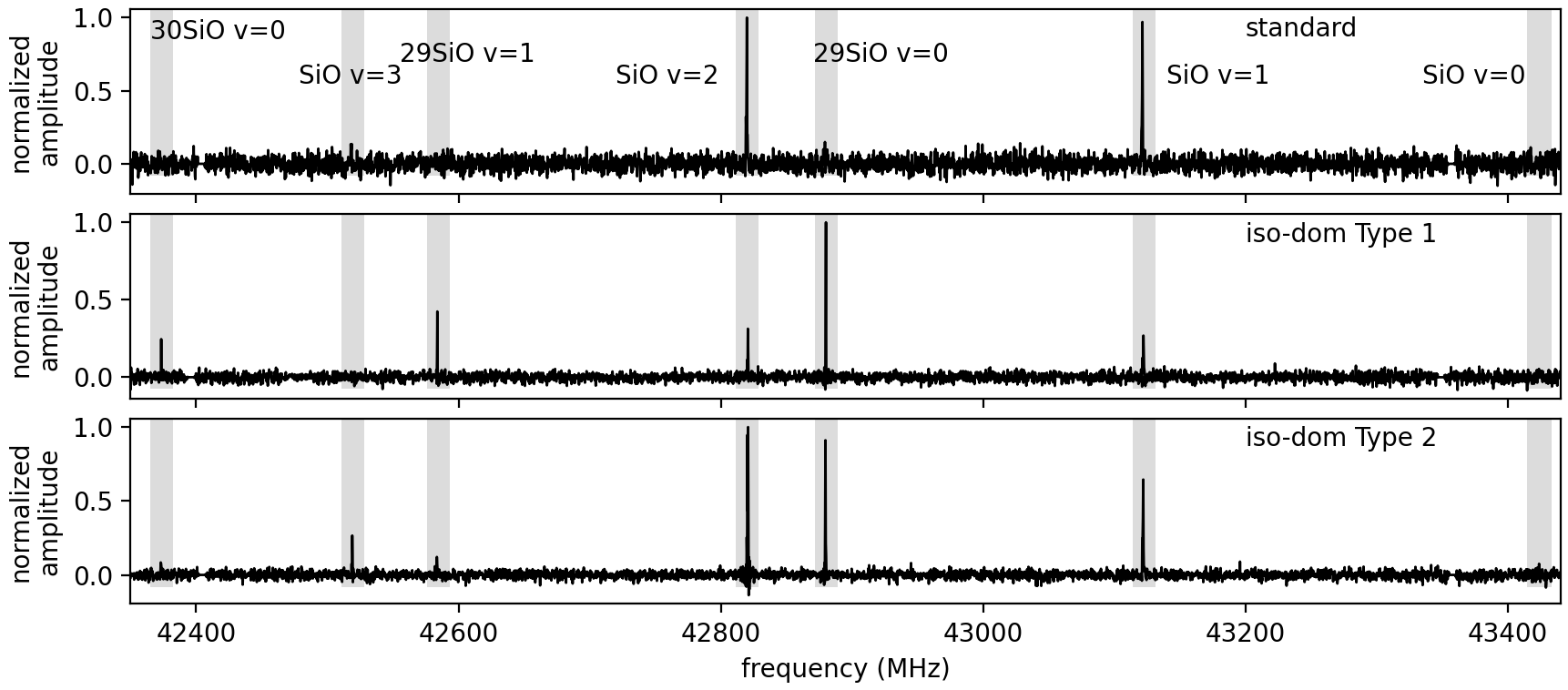}
    \caption{Three example SiO maser spectra. The top spectrum shows a typical detection with the SiO \textit{v}=1 and \textit{v}=2 being most prominent. The middle and bottom panels are examples of isotope-dominated (iso-dom) spectra, showing a Type 1 and Type 2 spectrum respectively (see text). All spectra are normalized and shifted into the rest frame. Frequencies where transitions are expected are shown in gray and labeled in the top panel. }
    \label{fig:threetypes}
\end{figure*}

There are two main categories of spectra which show bright isotopologue lines: one where the \SiOa\ \textit{v}=0 transition is the brightest, and a second where the SiO \textit{v}=2 is brightest followed closely by the \SiOa\ \textit{v}=0. Both show increased detection rates of all isotopologue lines, and share IR and period characteristics as discussed in Sect.\ref{sect:comm_chars}. Normalized example spectra of the two main subtypes of iso-dom sources are shown in Fig.\ref{fig:threetypes} along with an example of a more typical SiO maser signature. The division of these types is purely based on the observed maser spectrum, and it is unclear if there are physical differences in the stars or masers that fall into these categories, especially given the variable nature of maser emission. 

\subsection{Type 1 - \SiOa\ \textit{v}=0 dominates over both primary lines}

Of the 35 BAaDE sources, 22 show the \SiOa\ \textit{v}=0 as the brightest in the spectrum. We refer to this as spectrum Type 1. This includes sources where only \SiOa\ \textit{v}=0 is detected (4 sources), one source where only \SiOa\ and \SiOb\ lines are detected, and 17 where the SiO \textit{v}=0, 1, or 2 are detected but are weaker than the \SiOa\ \textit{v}=0. Eight of these 22 sources show \SiOa\ \textit{v}=1 emission, and another 8 show \SiOb\ \textit{v}=0 detections (not the same 8 though there is overlap). Thus, there is a much higher detection rate of these transitions than average BAaDE (0.2\% and 1.2\% for  \SiOa\ \textit{v}=1 and \SiOb\ \textit{v}=0, respectively, or 
0.2\% and 6.6\% within sources with primary lines over 9 times the detection threshold). We note that in the case of \SiOa\ \textit{v}=1, the detection rate is much higher in iso-dom spectra, but the detection rate does not rise in bright-primary-line sources; this illustrates how rare this line is and that it is truly associated with the iso-dom sources, not merely bright maser sources. Detections and maser flux densities can be found in Table \ref{tab:baadeoverview}, organized by subtype, then BAaDE name.

From \citet{rizzo21} and \citet{son25}, we also include S Cas as Type 1 using the arguments given in Sect.\,\ref{sec:discuss}.

\subsection{Type 2 - \SiOa\ \textit{v}=0 dominates over SiO \textit{v}=1}
Ten of the 35 BAaDE sources show SiO \textit{v}=2 as (marginally) brighter than the \SiOa\ \textit{v}=0, with the \SiOa\ \textit{v}=0 second brightest. This sample has significantly fainter or undetected (at our sensitivity) SiO \textit{v}=1 emission, thus the isotopologue line is brighter than the \textit{v}=1 primary line. These sources are labeled as spectrum Type 2. In 8 of these cases the spectra still resemble iso-dom spectra, with the \SiOa\ emission being comparable in flux density to the SiO \textit{v}=2; these sources often have detections of other \SiOa\ and \SiOb\ lines. Two sources (ad3a-09232 and ad3a-09257) however, look mostly like normal 43 GHz spectra with the SiO \textit{v}=1 line missing, i.e., the SiO \textit{v}=2 is much brighter than the \SiOa\ \textit{v}=0 and the SiO \textit{v}=1. Despite these two sources, among the Type 2 spectra the detection rate of \SiOa\ \textit{v}=1 (4/10) and \SiOb\ \textit{v}=0 (3/10) is quite high. We report all of these sources in our tables as the \SiOa\ \textit{v}=0 line is anomalously brighter than the SiO \textit{v}=1 in all cases, though we acknowledge there might be need for a further subdivision.  
The reverse case (weak or missing \textit{v}=2) is not common, but occurs in one case, see below. 

\subsection{Unique cases}
There are three especially unique cases: ad3a-10338, ad3a-12912, and ad3a-13671 marked as such in Tables \ref{tab:periods} and \ref{tab:baadeoverview}.

In the spectrum of ad3a-10338, the SiO \textit{v}=3 line is the brightest, followed by the SiO \textit{v}=2, the \SiOa\ \textit{v}=0, and finally the SiO \textit{v}=1. We have included it in this catalog as the \SiOa\ \textit{v}=0 is brighter than the SiO \textit{v}=1 (and comparable to the SiO \textit{v}=2); thus it is similar to Type 2, but we note that the SiO \textit{v}=3 emission is uniquely bright in this case such that it is unclear if this source is similar to the others. 

The spectrum of ad3a-12912 has two high signal-to-noise detections: the SiO \textit{v}=1 and the \SiOa\ \textit{v}=0; they are of comparable flux density. As noted this is the only iso-dom case of SiO \textit{v}=2 being prominently undetected from a spectrum where the \textit{v}=1 is found, the reverse being more typical.

Finally, the spectrum of ad3a-13671 shows only a detection of the SiO \textit{v}=1 and the \SiOa\ \textit{v}=1---a unique combination. As this spectrum shows no detection of the \SiOa\ \textit{v}=0 it was added to this catalog based on the detection of the \SiOa\ \textit{v}=1.

\begin{table}[h!]
    \caption{Variability properties}
    \label{tab:periods}
\centering
    \begin{tabular}{l|c|l|c h}
    name & date & period & phase &notes\\
     &  & (days) &  &\\
    \noalign{\hrule}\noalign{\smallskip}
    \textbf{Type 1} \\
    ad3a-00326 & 20160226 & 800\tablefootmark{a} & 0.63 &\\
    ad3a-01808\tablefootmark{b} & 20160227 & ... & ... & \\
    ad3a-04942 & 20130405 & ... & ... & \textit{v}=0 \\
    ad3a-06398 & 20130315 & 718.7 & 0.13 & bright \\
    ad3a-07500 & 20130303 & 850\tablefootmark{a} & 0.22 &  \\
    ad3a-07609 & 20130301 & ... & ... & ... \\
    ad3a-07856 & 20130302 & ...\tablefootmark{c} & ... & \\
    ad3a-08658 & 20160320 & 682.1\tablefootmark{d} & 0.75 & ...\\
    ad3a-10509\tablefootmark{b} & 20170320 & 597.9 & 0.94 & \\
    ad3a-10795 &20170318 & ... & ... & ... \\
    ad3a-10959 & 20170318 & ... & ... & ... \\
    ad3a-11129 & 20170325 & ... & ... & has \textit{v}=0 \\
    ad3a-11939\tablefootmark{e} & 20170310 & 760\tablefootmark{f} & ... & \\
    ad3a-12003 & 20160215 & ... & ... & ... \\
    ad3a-12022 & 20170310 & ... & ... & ...\\
    ad3a-12689 & 20170320 & 574.2\tablefootmark{d} & 0.60 & \\
    ad3a-12702 & 20170313 & 632.7 & 0.86 & ... \\
    ad3a-13236\tablefootmark{g} & 20170312 & 750\tablefootmark{a} & 0.12 & \\
    ad3a-13894 & 20170330 & 585.8 & 0.88 & ... \\
    ad3a-14355 & 20160415 & 608.7 & 0.34 & ... \\

    ce3a-00109\tablefootmark{e} & 20130303 & ... & ... & \\
    ce3a-00142\tablefootmark{e} & 20170327 & ... & ... & \\
    S Cas\tablefootmark{h} & 20120702 & 612.4 & 0.01 & \\
     & 20240317 &  &  0.92 & \\
    
    \noalign{\hrule}\noalign{\smallskip}
    \textbf{Type 2}\\
    ad3a-06591 & 20130312 & 525\tablefootmark{a} & 0.35 & \\
    ad3a-07522 & 20130302 & 577.7 & 0.96 & \\
    ad3a-08159 & 20160327 & ... & ... & \\
    ad3a-09232 & 20160316 & ... & ... & \\
    ad3a-09257 & 20170408 & ...\tablefootmark{c} & ...  & \\
    
    ad3a-10127 & 20170325 & 599.6\tablefootmark{d} & 0.16 & \\
    ad3a-11229 & 20170325 & ... & ... & ... \\
    ad3a-12626 & 20170313 & 664.7 & 0.41 & ... \\
    ad3a-13638 & 20170303 & 581.7\tablefootmark{d,i} & 0.83 & \\
    ce3a-00136 & 20170319 & ... & ... & ... \\
    
    \noalign{\hrule}\noalign{\smallskip}
    \textbf{Unique cases}\\
    ad3a-10338 & 20170320 & 574.2 & 0.93 & \\
    ad3a-12912 & 20170326 & 586.1 & 0.03 & \\
    ad3a-13671 & 20170309 & 516.1 & 0.06 & \\

    \end{tabular}
        \tablefoot{\\ 
        Variability properties of the thirty-five sources which show isotopologue-line dominated spectra at 43 GHz during the BAaDE survey plus S Cas.Periods are given in days, and phase zero or one indicates stellar (optical) maximum, with the minimum at phase 0.5.\\
    \tablefoottext{a}{period derived for this work from Gaia DR3 epoch photometry, $\pm$50 day precision}\\
    \tablefoottext{b}{reobserved, no-longer iso-dom \citep{lewis20b}}\\
    \tablefoottext{c}{no epoch photometry from Gaia}\\
    \tablefoottext{d}{OGLE data is sparse (less than 80 points)}\\
    \tablefoottext{e}{resobserved, iso-dom \citep{lewis20b}}\\
    \tablefoottext{f}{period from \cite{urago20}}\\
    \tablefoottext{g}{reobserved, nondetection \citep{lewis20b}} \\
    \tablefoottext{h}{maser observations by \cite{rizzo21,son25} both phases are shown}\\
    \tablefoottext{i}{less than one stellar cycle from OGLE}\\
    
    }
\end{table}

\begin{table*}[]
    \caption{Maser properties }
    \label{tab:baadeoverview}
    \centering
    \begin{tabular}{c|c|c|c|c|c|c|c}
    name  & \SiOb\ \textit{v}=0 & SiO \textit{v}=3 & \SiOa\ \textit{v}=1 & SiO \textit{v}=2& \SiOa\ \textit{v}=0 & SiO \textit{v}=1 & RMS \\
       & (Jy) & (Jy) & (Jy) & (Jy) & (Jy) & (Jy) & (mJy channel$^{-1}$) \\
    \noalign{\hrule}\noalign{\smallskip}
    \textbf{Type 1} \\
    ad3a-00326 & ... & ... & 0.18 & 0.12 & 0.32 & 0.13 & 22.9\\
    ad3a-01808 & ... & ... & ... & ... & 0.17 & ... & 18.1\\
    ad3a-04942 & ... & ... & ... & 0.10 & 0.25 & ... & 20.3\\
    ad3a-06398 & 0.17 & ... & 0.33 & 0.24 & 0.77 & 0.23 & 15.3\\
    ad3a-07500 &  0.31 & ... & 0.22 & 0.18 & 0.57 & 0.36 & 18.0\\
    ad3a-07609 &  ... & ... & ... & 0.12 & 0.14 & 0.09 & 15.5\\
    ad3a-07856 &  0.11 & ... & 0.07 & 0.11 & 0.26 & ... & 12.0\\
    ad3a-08658 &  0.11 & ... & ... & 0.15 & 0.34 & 0.29 & 16.3\\
    ad3a-10509 &  ... & ... & ... & ... & 0.15 & ... & 15.6\\
    ad3a-10795 &  0.06 & ... & 0.11 & ... & 0.22 & ... & 12.9\\
    ad3a-10959 &  0.07 & ... & 0.05 & 0.06 & 0.15 & 0.08 & 12.7\\
    ad3a-11129 &  ... & ... & ... & 0.08 & 0.28 & 0.13 & 12.7\\
    ad3a-11939 &  ... & ... & ... & ... & 0.16 & ... & 14.3\\
    ad3a-12003 &  0.07 & ... & 0.06 & 0.05 & 0.24 & 0.11 & 10.7\\
    ad3a-12022 & ... & ... & 0.07 & 0.08 & 0.15 & 0.07 & 13.0\\
    ad3a-12689 &  0.07 & 0.08 & ... & 0.10 & 0.26 & 0.18 & 15.9\\
    ad3a-12702 &  ... & ... & ... & 0.10 & 0.17 & 0.10 & 12.9 \\
    ad3a-13236 &  ... & ... & ... & ... & 0.11 & ... & 11.8\\
    ad3a-13894 &  ... & ... & ... & ... & 0.11 & 0.06 & 12.0\\
    ad3a-14355 &  ... & ... & ... & 0.17 & 0.47 & 0.32 & 17.7\\
    ce3a-00109 &  ... & ... & ... & ... & 0.25 & ... & 29.5\\
    ce3a-00142 &  ... & ... & ... & ... & 0.13 & ... & 15.2\\
    \noalign{\hrule}\noalign{\smallskip}
    \textbf{Type 2} \\
    ad3a-06591 &  ... & ... & ... & 0.26 & 0.20 & 0.09 & 34.6\\
    ad3a-07522 &  0.06 & 0.18 & 0.08 & 0.68 & 0.62 & 0.44 & 14.4\\
    ad3a-08159 &  ... & 0.12 & 0.33 & 0.89 & 0.42 & 0.14 & 13.8\\
    ad3a-09232 &  ... & 0.18 & ... & 1.11 & 0.15 & ... & 13.3\\
    ad3a-09257 &  ... & 0.24 & ... & 1.58 & 0.35 & 0.06 & 13.6\\
    ad3a-10127 &  ... & 0.06 & ... & 0.21 & 0.13 & 0.06 & 13.1\\
    ad3a-11229 &  0.08 & ... & ... & 0.26 & 0.12 & ... & 17.2\\
    ad3a-12626 & 0.08 & ... & 0.11 & 0.30 & 0.21 & 0.07 & 12.0\\
    ad3a-13638 &  ... & ... & 0.11 & 0.62 & 0.54 & 0.29 & 16.7\\
    ce3a-00136 &  ... & ... & ... & 0.32 & 0.17 & ... & 28.3\\
    \noalign{\hrule}\noalign{\smallskip}
    \textbf{Unique cases} \\
    ad3a-10338 &  ... & 0.20 & ... & 0.12 & 0.10 & 0.08 & 13.7\\
    ad3a-12912 &  ... & ... & ... & ... & 0.23 & 0.21 & 10.8\\
    ad3a-13671 &  ... & ... & 0.08 & ... & ... & 0.15 & 14.3\\
    \end{tabular}
    \tablefoot{Maser properties of the thirty-five sources which show isotopologue-line dominated spectra at 43 GHz during the BAaDE survey. Columns 2-7 show the flux densities of the maser transitions in a single peak channel (250 kHz width) in Jy, ordered from lowest frequency transition to highest. The last column shows the RMS noise in the spectrum in mJy/channel. Data for S Cas, which has a different spectral setup, can be obtained from \cite{rizzo21} (183 kHz channels) and \cite{son25} (15 kHz channels). The SiO \textit{v}=0 is not listed as it is not always a maser \citep{dike21}, and some detections are marginal. }
\end{table*}

\section{Common color and period trends among isotopologue-dominated sources}\label{sect:comm_chars}
We investigated commonalities among the iso-dom sources in terms of colors (from Gaia, 2MASS, MSX, WISE, AKARI, and IRAS photometry) and variability properties (from OGLE and Gaia data). All of the cross-matched source names are given in Table \ref{tab:otherobs}. We found periods for 20/36 sources, including S Cas, which showed an iso-dom spectrum in the observations of \cite{rizzo21} and \cite{son25}. Comments on the origin and reliability of the variability data are given in Table \ref{tab:periods}.

\begin{figure}
    \centering
    \includegraphics[width=0.9\linewidth]{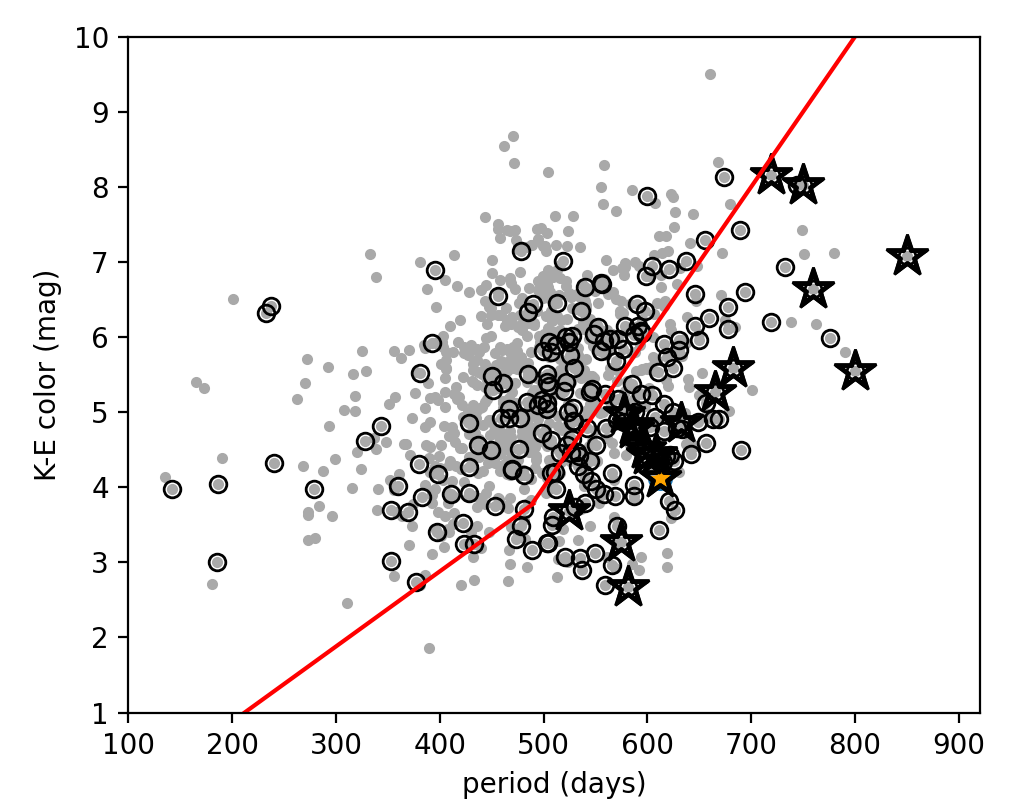}
    \includegraphics[width=0.9\linewidth]{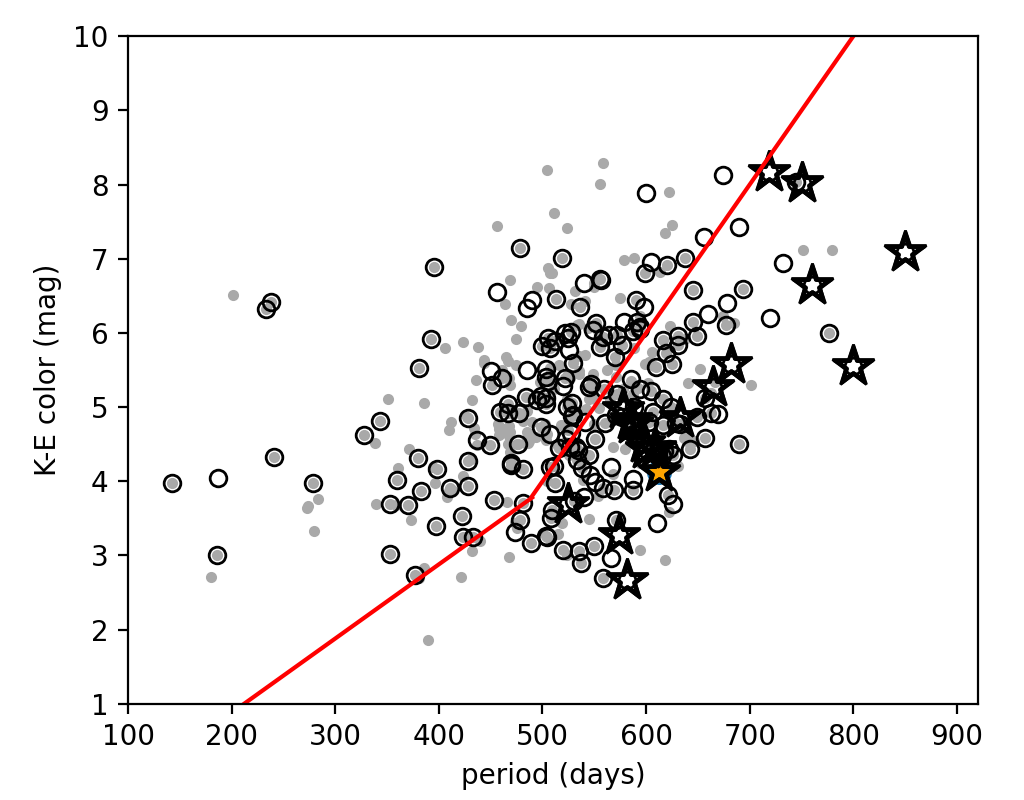}
    \caption{Color-period diagram using the K$-$E color (where K is the 2.2 micron magnitude from 2MASS and E is the 21$\mu$m magnitude from MSX). This plot is used as a diagnostic plot for TDU \citep{uttenthaler19} and the lower left portion of the red line (period$<$500 days) is their TDU divider based on short-period sources. Over 500 days we have modified the slope to 0.02 to empirically better separate the iso-dom sources. The top plot shows all BAaDE detections with available periods as gray dots, isotopologue detections as black rings, and the iso-dom sources as black stars, where S Cas is filled in orange. The bottom panel shows the BAaDE sources for which a primary line  detection is bright enough (9 times over the regular detection threshold) that we could expect an isotopologue detection as gray dots, and the other symbols are the same. Both representations show iso-dom sources, and isotopologue maser detections in general, are long-period for their color.}
    \label{fig:percol}
\end{figure}

When investigating near- and mid-IR colors, we find that the iso-dom sources tend to be long-period for their color, or from the other perspective, blue for their periods, as shown in Fig.\,\ref{fig:percol}.
These plots (specifically [K$_s$]$-$[22] versus period) have been used as diagnostic plots to separate AGB stars which have and have not undergone TDU \citep{uttenthaler19}. 
As all of our sources have MSX magnitudes we plot [K$_s$]$-$[E], where the MSX E is centered at 21$\mu$m \citep{egan03}. The exception is S Cas, which is outside of the MSX footprint and plotted by its [K$_s$]$-$[22] color. 
The stars which follow the bluer sequence, such as the sources with iso-dom spectra, are likely to have undergone TDU---an evolutionary period with deep convection zones where nucleosynthetically enriched material is brought into the CSE.
This division was established using a sample of much shorter-period sources, and \cite{uttenthaler19} point out that longer period sources in this region may be undergoing HBB as they are likely more massive. The dividing line between TDU and non-TDU sources from \cite{uttenthaler19}, based on and shown for sources with a period of up to 500 days, is empirically amended for sources with a period of over 500 days by a steeper line to better separate our iso-dom sources. Though the iso-dom sample shows the general trend of being long-period for their color, the \cite{uttenthaler19} division does not distinguish our iso-dom sample well, most likely because long-period sources were not included in their work. Conversely, we have no data for sources with periods shorter than 500 days, thus we include two distinct segments for the dividing line to emphasize that such empirical divisions depend on the range of periods included in the sample.

We note that all sources with isotopologue maser detections show this blue-for-period trend to some extent, not only the sources with spectra dominated by these masers. Sources which are blue for their period seem more likely to host \SiOa\ and \SiOb\ masers with both typical and atypical emission line ratios. Thus TDU, possibly in combination with HBB for the more massive ($\gtrsim$ 5 M$_{\odot}$) stars, may be necessary for isotopologue masers to be present in a spectrum, and may be a first step towards spectra with iso-dom line ratios. However, an additional (or several additional) criteria likely also need to be fulfilled to differentiate the fairly common occurrence of detecting isotopologue lines from the very rare iso-dom spectra.

A disproportionately high number of long-period sources show iso-dom spectra as shown by Fig.\,\ref{fig:percol}. The spectral Type 1 subtype is the only subtype with periods over 700 days, but otherwise the subtypes do not clump on color-period or color-color plots. 
However, the subtypes do show different behavior in the \textit{v}=3 line, which was not used to define the subtypes, showing that there is some predictive power in the division. 
Specifically, Type 1 show far fewer detections of the SiO \textit{v}=3 line than Type 2, and the unique source ad3a-10338 shows SiO \textit{v}=3 emission as brightest but otherwise resembles Type 2. 

The \textit{v}=3 line has been associated with higher-density environments \citep{desmurs14} and phases around stellar maximum \citep{cho96b}, but given the low-number statistics of this sample, we do not comment further. Without monitoring these sources we cannot determine if Type 1 and Type 2 are distinct objects or just a transient observational categorization.
Nevertheless, we present the division so that future detailed pumping models can aim to reproduce line ratios consistent with either/both subtypes.

The IRAS $[25]-[60]$ colors of the iso-dom sources are very red compared to other SiO maser stars. Using the \cite{vdvH88} IRAS color-color diagram as a guide, the iso-dom sources (only 8 with reliable IRAS matches) all lie in region VIb, whereas typical O-rich SiO maser sources are found mostly in region IIIa with some in II, IIIb, and VII. Region VIb is more typically associated with C-rich AGB stars, but sources in our sample are not C-rich stars because they display prominent SiO maser emission. It is possible the isotopologue sources are S-type stars as will be addressed in Sect.\ref{sect:tdu} and this could explain why their IR colors do not fit neatly into O-rich/C-rich divisions. 

\section{Discussion}\label{sec:discuss}
Using both our source list and the star S Cas, which has also shown an iso-dom spectrum \citep{rizzo21, son25}, but falls outside of the MSX footprint used in BAaDE, we discuss the possible effects of enhanced abundances due to TDU, stellar pulsation variability, and CSE turbulence velocity in terms of possible mechanisms for producing iso-dom spectra.

Several factors could lead to changes in the emission ratios of maser lines. Firstly, a relative increased abundance of \SiOa\ and \SiOb\ in the circumstellar envelope could contribute. Reversing the actual abundance ratio of SiO to \SiOa\ requires atypical nucleosynthetic processes in the core of the star, whereas subtler changes in this ratio are expected simply via the \textit{s}-process \citep{zinner06}. In either case, $^{29}$Si also needs to be transported into the CSE for these abundances to be reflected in the observed spectrum. This is achieved via dredge-up events where convection zones become deep enough to elevate nucleosynthetically enhanced material into the CSE. Thus both nucleosynthesis and dredge-up events contribute to any changes in abundance. 

A second option is for the \SiOa\ and \SiOb\ masers to be more efficiently pumped, meaning these transitions are more efficiently put into a population-inverted state. Changes in pumping efficiency between transitions that are very near in frequency can be caused by overlaps with other molecular transitions. 

Other options include more coherent velocities between the \SiOa\ and \SiOb\ molecules than the SiO molecule, a different pathlength for the \SiOa\ and \SiOb\ masers, or some kind of flaring in specific transitions; but these latter explanations require the various SiO species to have different kinematics, distributions, or temperatures. 
Both abundance- and pumping-related explanations are explored in more detail in the following sections.

\subsection{TDU}\label{sect:tdu}
The TDU phase brings elements from the stellar interior, including $^{29}$Si and $^{30}$Si produced via slow neutron capture, into the CSE via large convection cells. 
Thus it is not unreasonable to expect a slight increase in the CSE abundance of \SiOa\ and \SiOb\ in sources which have undergone this stage.

Overall, sources which show iso-dom spectra at 43 GHz tend to be blue for their periods; this period-color space is correlated with AGB stars which have undergone TDU. However, especially as compared to other sources which host isotopologue lines (with typical ratios), there is considerable overlap in period-color space. 
In order for a spectrum to display isotopologue lines there must be some \SiOa\ and \SiOb\ in the CSE; this is true whether or not the isotopologue lines dominate the spectrum.
It is clear that additional conditions need to be satisfied to differentiate sources which host detections of isotopologue masers with regular ratios from the iso-dom ones.

To our knowledge the only time a 43 GHz iso-dom spectrum has been observed outside of the BAaDE survey was documented in \cite{rizzo21} and later \cite{son25}, where the variable star S Cas showed brighter emission in the \SiOa\ \textit{v}=0 line than the SiO \textit{v}=1 or \textit{v}=2 lines in observations from 2012 and 2024, respectively. S Cas is a known S-type star \citep{stephenson84} meaning TDU has occurred, and there are \textit{s}-elements in the optical spectrum of S Cas \citep{keenan54}. 
The 86 GHz maser spectrum of S Cas yields a detection of the SiO \jtwo\ \textit{v}=2 maser line \citep{bujarrabal96, rizzo21}, which has also been associated with S-type stars \citep{bujarrabal96,soriaruiz04,michael19}.

S Cas also shows an asymmetry in its light curve characterized by an extra bump on the rising side of the curve (for example, after 2010 and after 2015 marked by arrows in Fig.\,\ref{fig:scas}). \cite{uttenthaler25} posit that these asymmetries are related to TDU. We see similar asymmetries in five of our sources with OGLE light curves (ad3a-14355, ad3a-13671, ad3a-12689, ad3a-10509, and ad3a-07522).

Although some signatures which aid in identifying S-type stars have been identified in the ALMA portion of the BAaDE survey at 86 GHz \citep{michael19}, no such distinction has yet been made for the VLA data at 43 GHz \citep{lewis20a} which has good coverage of the isotopologues.
With S Cas being the only iso-dom source known to be an S-type star, it is not definitive that the two are linked. However, as this fits the picture of a star which has undergone TDU---which physically motivates the presence of \SiOa\ and \SiOb\ in the CSE, and fits the period-color distribution of our sources---the association is notable. 

Stars which have undergone TDU or even which have already become S-stars seem good candidates to display isotopologue maser lines, though the complete reversal in line ratios in our iso-dom sources requires more explanation.  

\subsection{Variability}

One star, ad3a-01808, was observed to have an iso-dom spectrum in the BAaDE data from 2016, but \cite{lewis20b} reported the spectrum had typical ratios in 2020; albeit at this second epoch the flux densities of the masers were very low (lower than the noise in the first epoch, all lines $<$30 mJy). Thus it is clear that this signature can vary on the timescale of years, whereas the typical stellar cycle of an AGB star is of the order 1 to 2 years. Unfortunately, there is not a reliable light curve for this source in any of the major variability databases (OGLE, Zwicky, AAVSO) and it is thus unclear if the line ratio variability could be related to stellar phase. Still the variability on a few years timescale makes it highly unlikely that the line ratios in ad3a-01808 are linked to an isotopologue abundance anomaly where there is significantly more \SiOa\ in the envelope than SiO. ad3a-01808 showed a single detection of \SiOa\ \textit{v}=0 during the 2016 epoch and thus has a Type 1 iso-dom spectrum. The question as to whether this is a general behavior of these sources remains. 

Similarly, the star ad3a-10509 was iso-dom in the 2017 BAaDE data (single-line detection of \SiOa\ \textit{v}=0, Type 1 iso-dom spectrum), but showed only a detection of (likely thermal) SiO \textit{v}=0 in follow-up observations in 2020 \citep{lewis20b}. In the follow-up observations the phase was $\sim$0.70, compared to the $\sim$0.94 phase of the BAaDE detection. Thus the iso-dom behavior was observed closer to stellar maximum than the thermal SiO detection. 

\begin{figure*}
    \sidecaption{}
    \centering
    \includegraphics[width=12.0cm]{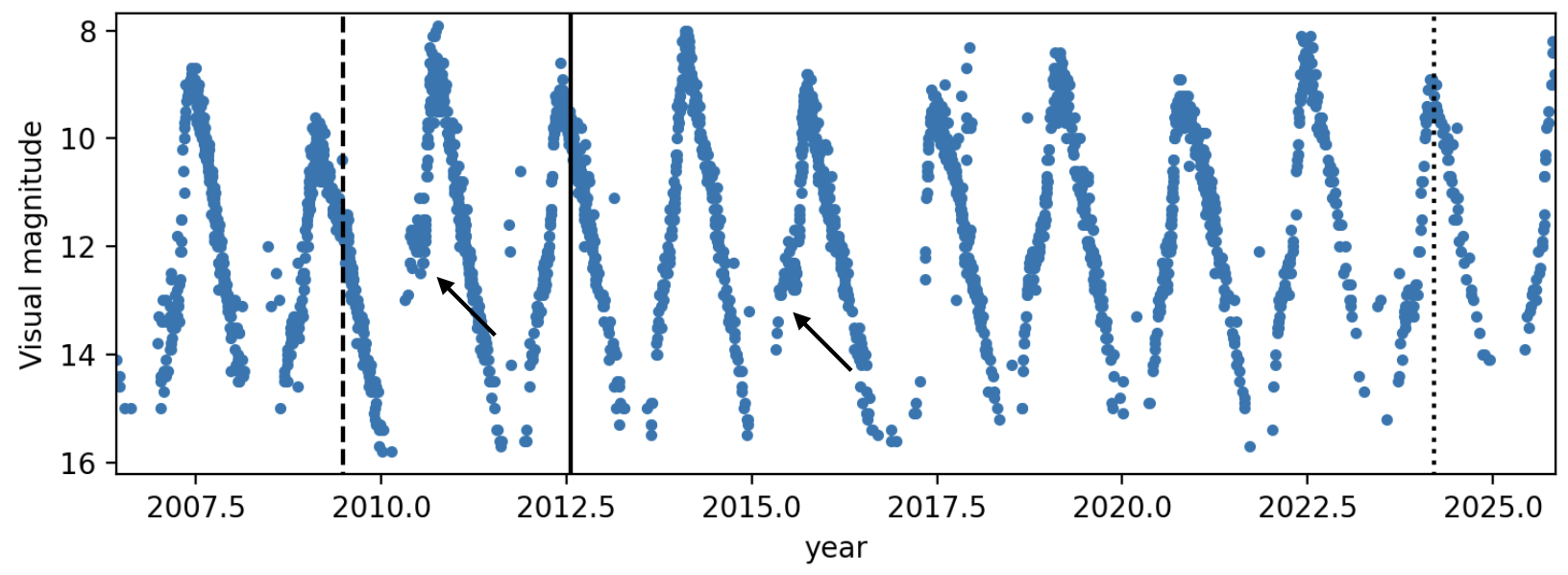}
    \caption{AAVSO light curve of S Cas, with SiO observation dates marked. The observations of \cite{cho12} (date marked as dashed line) showed no detection of the \jone\ lines, while the observations of \cite{rizzo21} (date marked as a solid line) and \cite{son25} (dotted line) showed an iso-dom \jone\ spectrum. Two examples of bumps on the rising side of the light curve are pointed out by small arrows (see text).
    \vspace{0.6cm}}
    \label{fig:scas}
\end{figure*}

\begin{figure}
    \centering
    \includegraphics[width=0.95\linewidth]{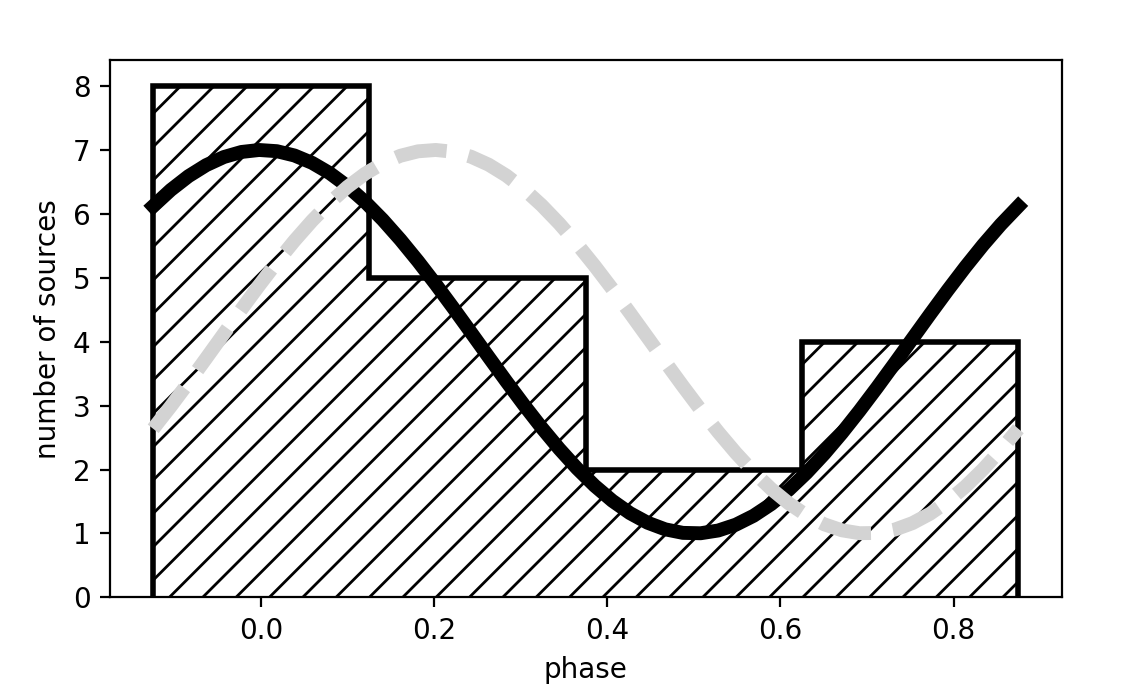}
    \caption{Optical phases of the maser observations of the iso-dom sources. To guide the eye, a sinusoidal model light curve with minimum light at a phase of 0.5 and maximum at 0, representative of the optical light curve, is plotted in black; while a light curve shifted by 0.2 in phase, representative of a primary SiO maser light curve \citep{alcolea99}, is shown in dotted gray. }
    \label{fig:phasehist}
\end{figure}

S Cas has also shown variable maser line behavior. It was observed in the \jone\ lines by \cite{rizzo21} for two epochs (02 July and 19 July 2012). 
In those observations, \SiOa\ \textit{v}=0 was the brightest detected line, thus the spectrum is Type 1 iso-dom.
In 2023 it was observed by \cite{son25} in the \jone\ and also showed a Type 1 iso-dom spectrum.
S Cas has also been observed at 43 GHz by \cite{cho12} and \cite{spencer81}. 
Both these 43 GHz observations were nondetections of the SiO \textit{v}=1 and 2 lines, albeit with high upper limits of $\sim$0.2 and 13 Jy for the \cite{cho12} and \cite{spencer81} observations, respectively. The \jone\ masers detected in 2012 (SiO \textit{v}=1 and \textit{v}=2, and \SiOa\ \textit{v}=0 and \textit{v}=1) showed flux densities between 2 and 8 Jy and would have been detected in the \cite{cho12} observations if they were not variable. Similarly, the 2023 detections of SiO \textit{v}=2 and \SiOa\ \textit{v}=0 showed flux densities between 2 and 4 Jy and also would have been detected in the \cite{cho12} observations if they were not variable.
Figure \ref{fig:scas} shows the dates of the \cite{cho12}, \cite{rizzo21}, and \cite{son25} observations on a visual light curve from AAVSO \citep{watson06}. This shows that the observations were taken at relatively similar stellar phases, with the nondetection taking place slightly after visual maximum and both detections essentially at visual maximum, when stellar pulsation has reached its smallest physical extent. Further, S Cas was observed to be iso-dom in two observations over 10 years apart. Given that there are reports of a $\sim$ 10-20\% lag between primary SiO maser emission and optical or IR light curves \citep{alcolea99, pardo04, gonidakis13}, it is possible that for the \SiOa, the nondetection was observed closer to maser maximum than the detection; however we emphasize that the stellar phases are only marginally different.

In Table \ref{tab:periods}, we report the optical light curve phase at which the iso-dom spectrum was observed for each source with an OGLE or Gaia light curve.
Phases were determined by fitting the light curve data with a sine function. For the OGLE data we used the pre-determined amplitude, period, and average from the OGLE catalog, only fitting for a shift in the fit. We note that although OGLE data usually has a very long time-baseline, the redder sources we discuss here often have more sparse coverage---sometimes less than a complete cycle. Sources with sparse or limited-time-coverage OGLE data are marked in Table \ref{tab:periods}. Phases and periods derived from Gaia are also marked because the coverage in this data is also sparse given the $\sim$ 2 year periods of most of these sources. Especially the phase of ad3a-13236 is essentially unconstrained given the Gaia coverage, but our best guess is given. Sources without a note in Table \ref{tab:periods} have phases taken from OGLE data with sufficient coverage. All of the phases, including those from well-sampled data, warrant caution because of the non-symmetric nature of Mira variability which contributes to uncertainties in fitting and obtaining phases. Thus we caution that our listed phases are imprecise, but the statistical distribution of the sample may give insight.  
There is a preference towards observations at phases near optical maximum (phases below 0.25 or above 0.75) as shown in Figure \ref{fig:phasehist}. 
This could be driven by an observational bias towards stars in the brightest part of their cycles, as all SiO masers are also brighter and preferentially detected in these phases \citep{lewis24}. 
Type 2 sources especially clump around phases of 0.4 and 0.9---essentially minimum and maximum---but here we are working with only 5 sources with phases available. 
The unclassified, unique cases all have phases near maximum, though this being the subtype with miscellaneous spectra makes this hard to interpret. Better sampled light curve data is required to address the phases of these detections more robustly; this would require multi-year monitoring.

We also note that if we uniformly apply a 0.2 phase shift, as can be expected between the optical light curve and primary SiO maser variability, we retain a majority of sources in the brighter part of their cycles, but shift more sources into the rising part of their cycle. It may be the case that our iso-dom sources are more associated with optical maximum than with expected primary SiO maser maximum.
However, sample size, bias towards stellar maximum, and imprecision and inaccuracies in stellar phase calculations make it difficult to draw any quantitative conclusions here. We emphasize the need for both visual and maser monitoring.

In short, from current observations it is clear that iso-dom spectra are not a persistent feature over year timescales, though we cannot resolutely link this behavior to maximum stellar phase.

\subsection{Line-overlaps and turbulence velocity}
We explore the case where a maser-pumping mechanism, in addition to a slight change in abundance caused by TDU, is responsible for the atypical line ratios. As line overlaps are responsible for the strength and even existence of many isotopologue masers \citep{gac97}, we explore these overlaps as a possible cause.
\cite{gac97} showed that the \SiOa\ \textit{v}=0 and \textit{v}=1 can both be enhanced by the same overlap: \SiOa\ (\textit{J}=1, \textit{v}=1) $\rightarrow$ (\textit{J}=0, \textit{v}=0) and SiO (\textit{J}=2, \textit{v}=4) $\rightarrow$ (\textit{J}=1, \textit{v}=3), where the difference in the velocities of the overlapping transitions is 0.7
km s$^{-1}$. This lends an explanation to the close link between the two maser lines in our data. Further, they show that \SiOb\ \textit{v}=0 is produced by an overlap: \SiOb\
(\textit{J}=1, \textit{v}=1) $\rightarrow$ (\textit{J}=0, \textit{v}=0) and \SiOa\ (\textit{J}=1, \textit{v}=3) $\rightarrow$ (\textit{J}=0, \textit{v}=4), with a similarly small difference in velocity (0.5 km s $^{-1}$). In their model all three maser transitions are enhanced when turbulence velocity is low ($v_{tur}<$ 1 km s$^{-1}$), as other overlaps with larger velocity differences can cause competition at larger turbulence velocity. Thus very low turbulence velocity ($v_{tur}\lesssim$ 1 km s$^{-1}$), where these overlaps are relevant compared  to other overlaps, could be the condition that causes iso-dom spectra. 

The velocity spread may be reduced when a star transitions from expanding to contracting and vice versa, i.e., when the star is near maximum and minimum light, because at these moments pulsating material is changing direction and should have a velocity of 0 km s$^{-1}$ with respect to the star. Due to geometry, during optical maximum the stellar emission and the density in the CSE are higher than at optical minimum which should be more conducive to masers regardless of whether they are collisionally or radiatively pumped. This aligns with iso-dom spectra being observed at optical maximum, though there may be observational biases. If this low-velocity-spread transition is a set fraction of the pulsation period, then longer period stars will spend more time in this transition; thus, we may catch more long-period sources in this state, which is again consistent with our sample. 

Our conjecture is that stars which have undergone TDU, perhaps to the point of being S-type AGB stars, can show iso-dom spectra in cases where the masing gas has a low turbulence velocity. Thus the role of maser pumping and turbulence velocity would be responsible for the quick variations in the spectra. 

\subsection{Abundance in the Galaxy}

In \cite{para95} it is suggested that strong \SiOa\ signatures could be used to identify red (super) giants with neutron star cores, also known as Thorne-Żytkow objects (TŻO). Based on the variability and number of the objects described here, dominance of \SiOa\ maser lines cannot be used to unambiguously detect TŻOs.

Estimates of TŻO formation rate range from $\sim2\times10^{-5} - 3\times10^{-4}$ objects per year in the Milky Way \citep{podsiadlowski95, hutilukejiang18, romagnolo25}, which given their short lifetime ($\sim$10$^5$ years), allows for $\sim$30 TŻOs in the Milky Way on the high end. It is therefore highly unlikely that the VLA portion BAaDE survey, which covered $\sim$19000 AGB stars from specific color ranges of the MSX catalog, detected 35 TŻOs -- especially given that we have shown that this maser signature can turn off. Further, iso-dom sources are not clearly associated with molecular clouds in the Galactic plane. Such an association is seen clearly for other young objects in the BAaDE sample (young stellar objects, see \cite{lewis20a}), and we would expect to see this association again if iso-dom objects were young red supergiants.

On the other hand, S-type stars are much more common in the Milky Way, with several thousand identified \citep{stephenson84, stephenson90, chen22, chen23}. If we simply use \jtwo\ SiO \textit{v}=2 detections as a proxy for S-type stars in the ALMA portion of the BAaDE survey, we can expect 4\% of BAaDE targets to be S-type \citep{michael19}. This yields a rough estimate of $\sim$ 750 S-type sources in the VLA portion of the survey. 
By this estimate only a small fraction of S-type stars in BAaDE (5\%) would be displaying isotopologue-dominated spectra. Thus extra criteria beyond S-type status would certainly be required for iso-dom spectra. As discussed above, these extra criteria are likely time variable and related to maser pumping; we put forward very low turbulence velocity as a possible explanation.

\section{Conclusions}

We report 36 maser sources with anomalous 43 GHz spectra (35 from the BAaDE catalog and observations of the S-type star S Cas \citep{rizzo21,son25}), with high relative brightnesses of \SiOa\ \textit{v}=0 and 1 and \SiOb\ \textit{v}=0 lines. These iso-dom spectra can be divided into two main categories: those where the \SiOa\ \textit{v}=0 is the brightest in the spectrum, and those where the \SiOa\ \textit{v}=0 is second brightest after the SiO \textit{v}=2. 

All of these sources have relatively long periods for their K$-$E color, which has been used as a diagnostic for sources which have undergone TDU, though our sample averages much longer periods than the stars typically used for this diagnostic. S Cas is a known S-type star and has undergone TDU events. Thus TDU or even  S-type status likely boosts the amount of $^{29}$Si and $^{30}$Si in the CSE---this is supported by modeling \citep{zinner06}---and may be one of the baseline criteria for these maser anomalies. 

Further, the three main \textit{J}=1---0 \SiOa\ and \SiOb\ lines probed by BAaDE are produced by line overlaps which are very close in velocity (0.7 and 0.5 km s$^{-1}$ for the overlaps which produce the \SiOa\ and \SiOb\ lines respectively) according to the model in \cite{gac97}. We therefore posit that low turbulence velocity, where there is not enough velocity spread for other overlaps to be present, aids in producing the atypical spectra. In this scheme both TDU and low turbulence velocity are required for such spectra; with TDU yielding a slight increase in \SiOa\ and \SiOb\, and the narrow turbulence velocity requirements provide favorable maser pumping conditions explaining the variability and rarity of these spectral features. Although our data could suggest iso-dom masers occur near the stellar maximum, due to bias towards observations at stellar maximum and the small sample size, we cannot firmly draw this conclusion. We reject that these iso-dom maser sources are TŻOs in general, but cannot exclude individual exceptions. 
Monitoring campaigns and more detailed maser pumping modeling will be required to more fully understand this signature.

\begin{acknowledgements}
      
      We gratefully acknowledge the contributions of the AAVSO observer community, whose photometric data and metadata resources were used in this study and made available through the AAVSO’s scientific archives. 
      
      This research has made use of the SIMBAD database, operated at CDS, Strasbourg, France.

RB acknowledges support for this work provided by the National Science Foundation (NSF) through the Grote Reber Fellowship Program administered by Associated Universities, Inc./National Radio Astronomy Observatory (AUI/NRAO).
      The National Radio Astronomy Observatory is a facility of
the U.S.\ National Science Foundation operated under cooperative
agreement by Associated Universities, Inc.
      
     This work has made use of data from the European Space Agency (ESA) mission
{\it Gaia} (\url{https://www.cosmos.esa.int/gaia}), processed by the {\it Gaia}
Data Processing and Analysis Consortium (DPAC,
\url{https://www.cosmos.esa.int/web/gaia/dpac/consortium}). Funding for the DPAC
has been provided by national institutions, in particular the institutions
participating in the {\it Gaia} Multilateral Agreement.

This research made use of data products from the Midcourse
Space Experiment. Processing of the data was funded by the
Ballistic Missile Defense Organization with additional support
from NASA Office of Space Science.

This publication makes use of data products from the Two
Micron All Sky Survey, which is a joint project of the
University of Massachusetts and the Infrared Processing and
Analysis Center/California Institute of Technology, funded by
the National Aeronautics and Space Administration and the
National Science Foundation.
\end{acknowledgements}

%
   \bibliographystyle{aa} 
   \bibliography{thesisbib_copyplus.bib} 
%

\begin{appendix}
\section{Cross-matched sources}
Table \ref{tab:otherobs} presents the names of the isotopologue-dominated sources in the surveys and databases used in this work. 

\begin{sidewaystable*}[]
     \caption{Thirty-six sources which show isotopologue-line dominated spectra at 43 GHz. Thirty-five spectra are from the BAaDE survey, and S Cas was observed by \cite{rizzo21} and \cite{son25}.}
     \centering
     \begin{tabular}{c|h|c|c|c|c|c|c}
         BAaDE & IRAS & 2MASS & MSX & WISE & AKARI & Gaia DR3 & OGLE\\
          ad3a-00326  & 17252$-$3359 & J17283584$-$3401237 &  G353.7609+00.3279 & J172835.84$-$340123.4 & J1728358$-$340122 & 5975965994709270912 & ... \\
          ad3a-01808  & 17398$-$3016& J17430552$-$3018014 & G358.5493$-$00.2454 & J174305.51$-$301801.3 & J1743053$-$301802 & ... & ... \\
         ad3a-04942  & 17455$-$2725 & J17484093$-$2725587 & G 001.6346+00.2021 & J174840.85$-$272558.5 & ... & ... & ... \\
         ad3a-06398  & 17538$-$2345 & J17565328$-$2345293 & G005.7404+00.4784 & J175653.27$-$234529.1 & ... & 4069289991783534208 & BLG-LPV-255142 \\ 
         ad3a-07522  & 17540$-$2125 & J17570529$-$2126162 & G007.7708+01.6021 & J175705.38$-$212615.7 & J1757052$-$212616 & 4070409367676692480 & BLG-LPV-255255 \\
         ad3a-07500  & 17581$-$2130 & J18010916$-$2130204 & G008.1832+00.7486 & J180109.16$-$213019.0 & J1801091$-$213020 & 4070732486645625984 & ... \\
         ce3a-00109  & 17597$-$2219 & J18024826$-$2219523 & G007.6551+00.0075 & J180248.31$-$221952.2 & J1802482$-$221950 & ... & ... \\        
         ad3a-07609  & ... & J18062791$-$2110206 & G009.0819$-$00.1653 & ... & ... & ... & ... \\
         ad3a-06591  & 18056$-$2319 & J18084255$-$2318387 & G007.4652$-$01.6583 & ... & J1808425$-$231838 & 4066527507517542016 & ... \\
         ad3a-08159  & 18068$-$1910??? & J18094795$-$1909479 & G011.2183+00.1265 & ... & ... & ... & ... \\
         ad3a-07856  & 18075$-$2019 & J18103119$-$2018208 & G010.2997$-$00.5738 & ... & J1810313$-$201821 & 4094220425815918592 & ... \\
         ad3a-10127  & 18146$-$1146 & J18172538$-$1145416 & G018.5971+02.0568 & ... & J1817254$-$114542 & 4153877762068657024 & BLG-LPV-261297 \\
         ad3a-08658  & 18159$-$1719 & J18185365$-$1718281 & G013.8799$-$00.8832 & ... & ... & 4097310117901098880 & BLG-LPV-261717 \\
         ad3a-09257  & ... & J18210454$-$1502362 & G016.1235$-$00.2764 & J182104.56$-$150236.6 & ... & 4098232097056488832 & ... \\
         ad3a-09232  & 18186$-$1510 & J18213022$-$1508394 & G016.0831$-$00.4147 & J182130.25$-$150839.4 & J1821302$-$150840 & ... & ... \\
         ce3a-00136  & 18234$-$1306 & J18261734$-$1304162 & G018.4583$-$00.4689 & J182617.17$-$130415.9 & J1826173$-$130415 & ... & ... \\
         ad3a-10509  & ... & J18304074$-$1044261 & G021.0209$-$00.3371 & J183040.77$-$104425.6 & ... & 4154959063049825280 & GD-LPV-018381 \\ 
         ad3a-10795  & 18282$-$1014 & J18310234$-$1012129 & G021.5380$-$00.1671 & J183102.31$-$101213.1 & J1831023$-$101213 & ... & ... \\
         ad3a-10959  & ... & J18311953$-$0945273 & G021.9659$-$00.0234 & ... & ... & ... & ... \\
         ad3a-11129  & 18286$-$0910 & J18312512$-$0908158 & G022.5262+00.2428 & J183125.17$-$090815.6 & ... & ... & ... \\
          ad3a-11229  & 18293$-$0840 & J18320576$-$0837487 & G023.0540+00.3291 & J183205.83$-$083749.9 & J1832058$-$083748 & ... & ... \\
         ce3a-00142  & 18312$-$1209 & J18340225$-$1206501 & G020.1821$-$01.7012 & J183402.22$-$120650.0 & ... & ... & ... \\
         ad3a-12689 & 8325$-$1107 & J18352267$-$1105039 & G028.4024$-$00.3810 & J184431.31$-$041216.3 & J1835226$-$110503 & 4154734934496598784 & BLG-LPV-265199 \\
         ad3a-12022  & ... & J18370853$-$0631120 & G025.5029+00.1913 & ... & ... & ... & ... \\
         ad3a-12003  & 18346$-$0638 & J18371806$-$0635323 & G025.4569+00.1233 & J183718.11$-$063533.6 & J1837180$-$063531 & ... & ...\\
         ad3a-12912  & 18366$-$0343 & J18391616$-$0340489 & G028.2694+01.0236 & J183916.14$-$034048.8 & J1839161$-$034046 & 4257026486964917504 & GD-LPV-019446 \\
         ad3a-13236  & 18397$-$0254 & J18422489$-$0251580 & G029.3527+00.6984 & J184224.90$-$025158.1 & J1842249$-$025157 & 4259422632035704064 & ... \\
         ad3a-11939  & 18404$-$0645 & J18430720$-$0642401 & G026.0130$-$01.2156 & J184307.30$-$064239.5 & ... & ... & ... \\
         ad3a-12626  & 18412$-$0425 & J18435658$-$0422007 & G028.1923$-$00.3269 & ... & ... & 4258226432095597440 & GD-LPV-020563 \\
         ad3a-10338  & 18325$-$1107 & J18443135$-$0412165 & G021.2481$-$01.5191 & J184431.36$-$041216.8 & ... & ... & BLG-LPV-265199 \\
         ad3a-13638  & 18422$-$0054 & J18445050$-$0051453 & G031.4123+01.0743 & ... & J1844505$-$005145 & 4260285370672161152 & GD-LPV-020766 \\        
         ad3a-13671  & 18448$-$0047 & J18472873$-$0044178 & G031.8236+00.5443 & J184728.74$-$004417.8 & J1847287$-$004416 & 4260121028047203712 & GD-LPV-021278 \\
         ad3a-12702  & 18493$-$0413 & J18520259$-$0409548 & G029.2924$-$02.0324 & J185202.65$-$040953.5 & J1852026$-$040955 & 4255259567442015616 & GD-LPV-022138 \\
         ad3a-13894  & ... & J18551237$-$0002583 & G033.3171$-$00.8608 & ... & ... & 4266223966111666560 & GD-LPV-022724 \\
         ad3a-14355  & 18576+0140 & J19001398+0145081 & G035.4945$-$01.1567 & J190013.99+014508.2 & J1900140+014508 & 4268376882887707008 & GD-LPV-023447 \\
         S Cas\tablefootmark{a} / NA& 01159+7220 & J01194198+7236407 & NA & J011941.77+723642.8 & ... & 535066514639376896 & NA \\
    \end{tabular}
     \tablefoot{
     \tablefoottext{a}{not observed by BAaDE, and the only previously reported instance of this behavior at 43 GHz to our knowledge. NA: Not in the survey footprint.}}
     \label{tab:otherobs}
 \end{sidewaystable*}

\section{Velocity profiles}
\begin{figure*}
\caption{Velocity profiles of the isotopologue-dominated sources}
    \centering
    \includegraphics[width=0.33\textwidth]{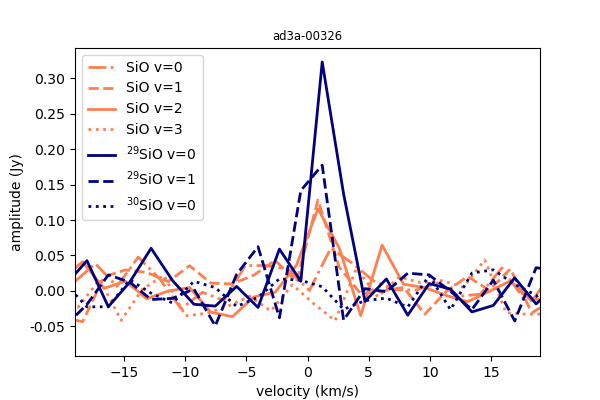}
    \includegraphics[width=0.33\textwidth]{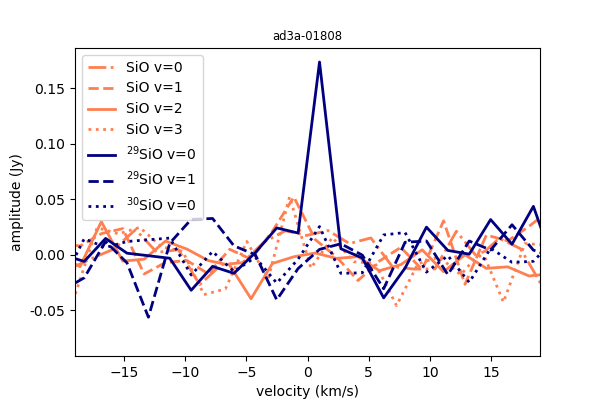}
    \includegraphics[width=0.33\textwidth]{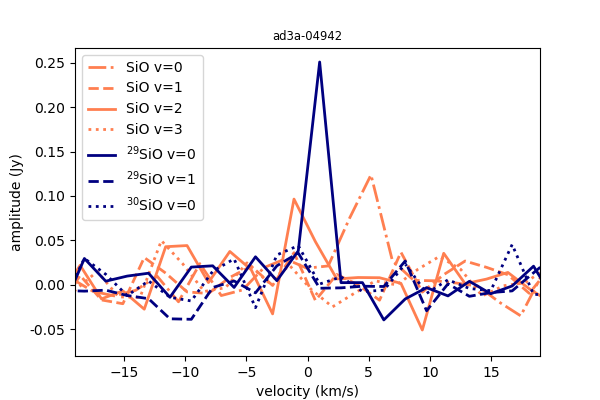}
    \includegraphics[width=0.33\textwidth]{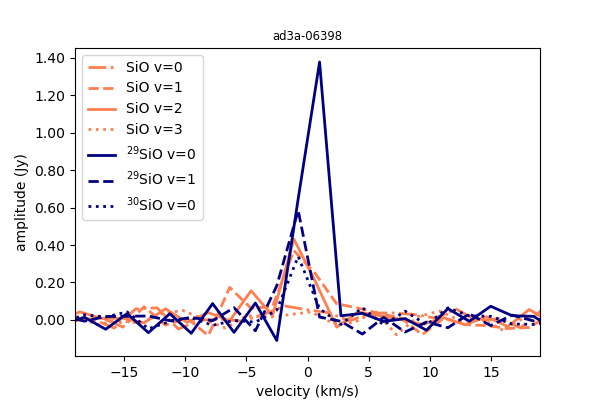}
    \includegraphics[width=0.33\textwidth]{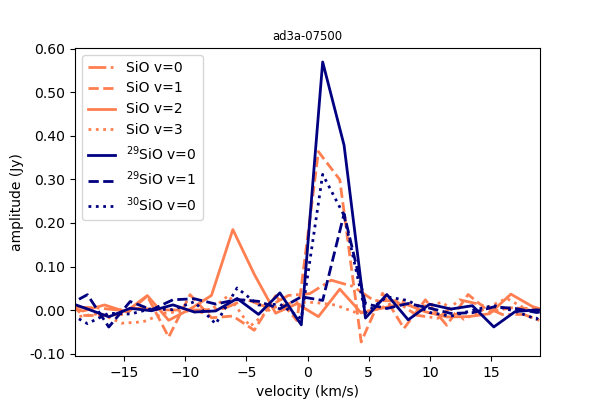}
    \includegraphics[width=0.33\textwidth]{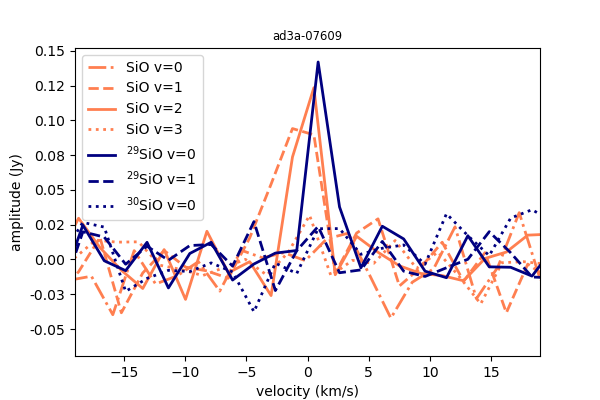}
    \includegraphics[width=0.33\textwidth]{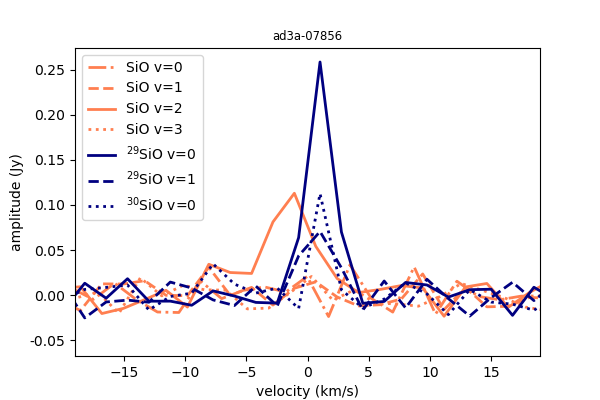}
    \includegraphics[width=0.33\textwidth]{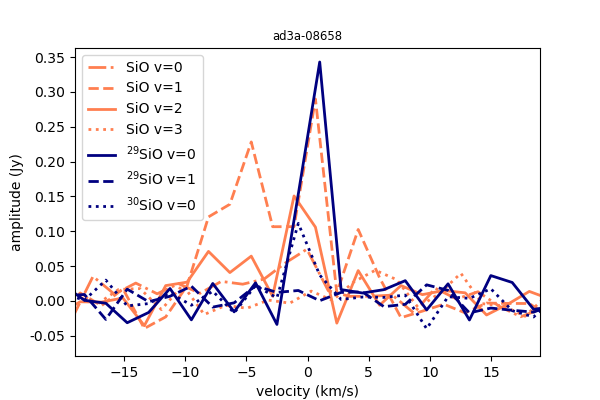}
    \includegraphics[width=0.33\textwidth]{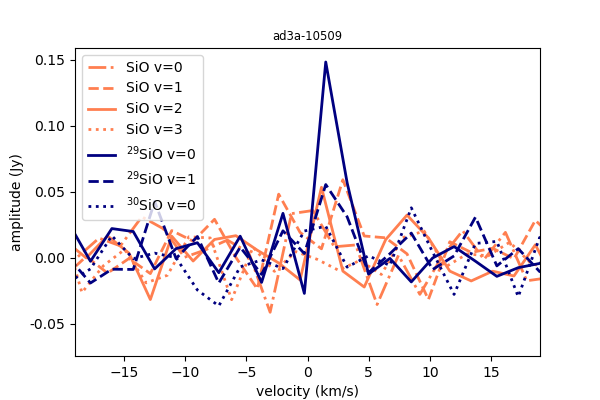}
    \includegraphics[width=0.33\textwidth]{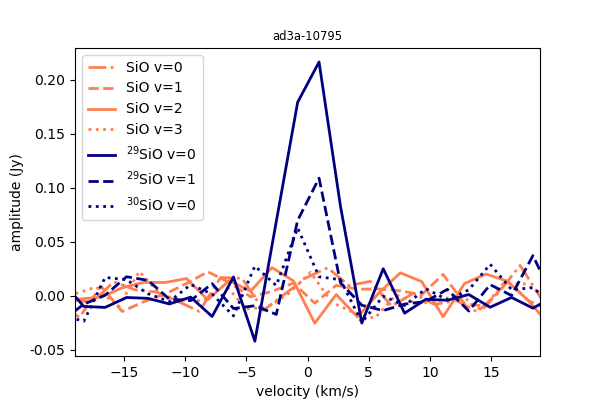}
    \includegraphics[width=0.33\textwidth]{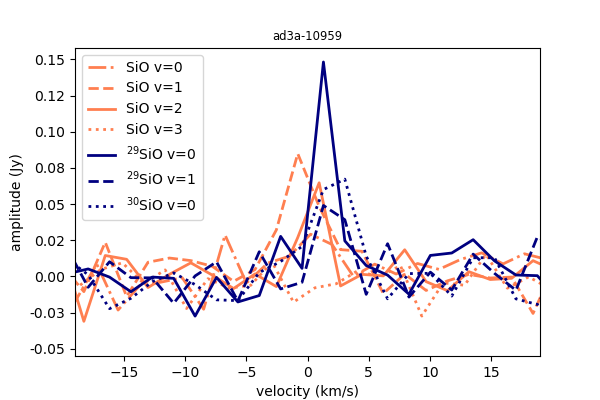}
    \includegraphics[width=0.33\textwidth]{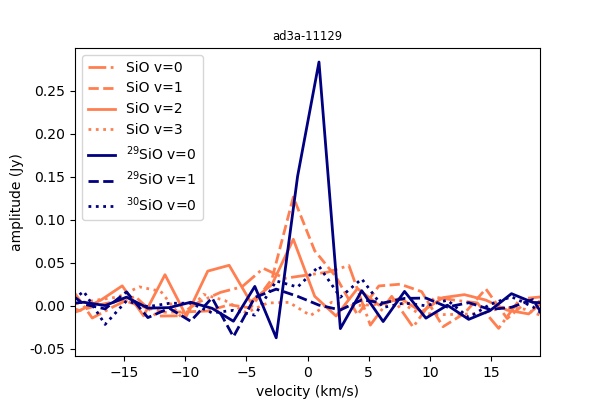}
    \includegraphics[width=0.33\textwidth]{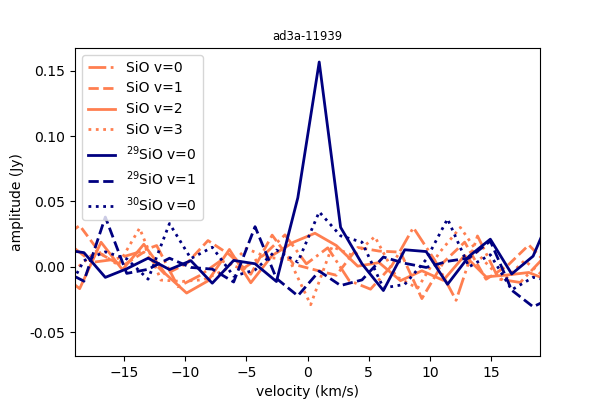}
    \includegraphics[width=0.33\textwidth]{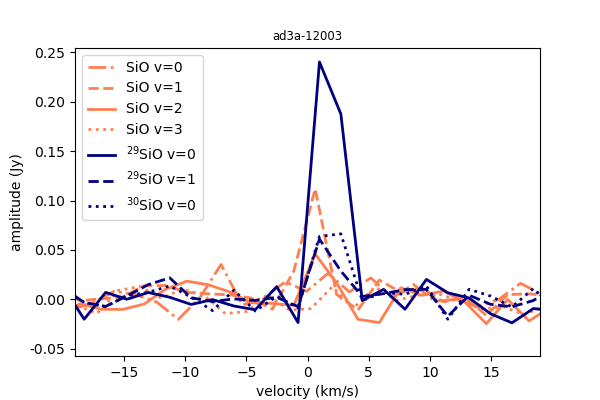}
    \includegraphics[width=0.33\textwidth]{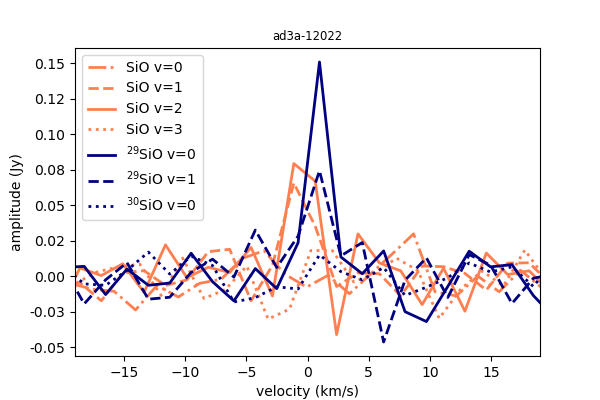}
    \includegraphics[width=0.33\textwidth]{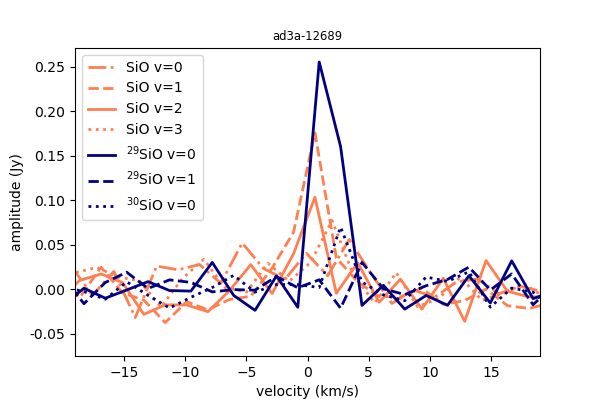}
    \includegraphics[width=0.33\textwidth]{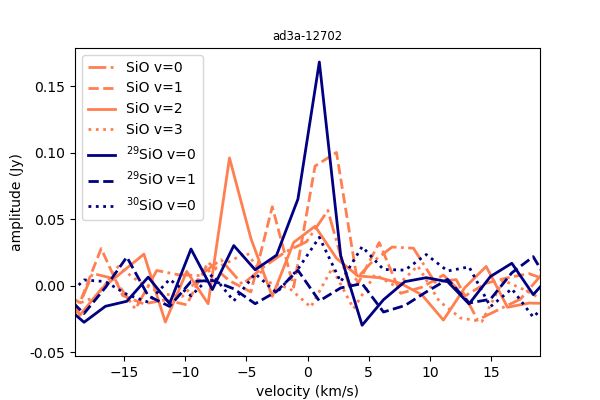}
    \includegraphics[width=0.33\textwidth]{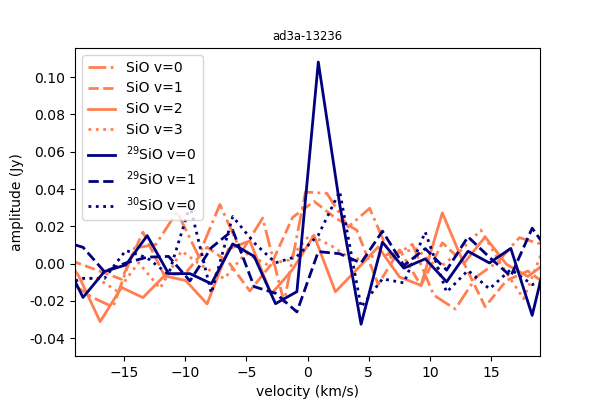}
   \label{fig:profspec}
\end{figure*}

Figure \ref{fig:profspec} presents the velocity profile of each of the 35 isotopologue-dominated sources from the BAaDE catalog, ordered by Type and BAaDE name (as in Table \ref{tab:baadeoverview}). For each isotopologue transition listed in Table \ref{tab:lines} the spectral profile is shifted to the systemic line-of-sight (LSR) velocity of the source. 


\addtocounter{figure}{-1}
\begin{figure*}
    \centering
    \includegraphics[width=0.33\textwidth]{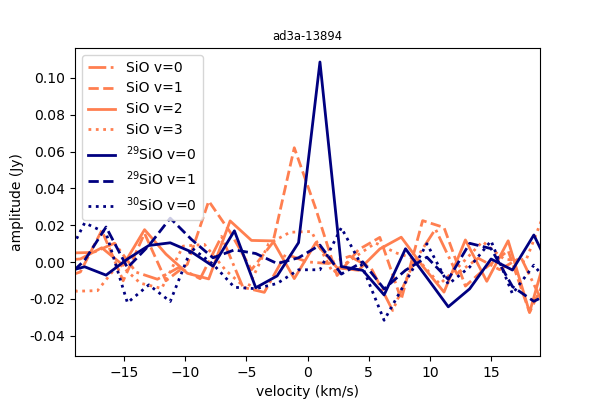}
    \includegraphics[width=0.33\textwidth]{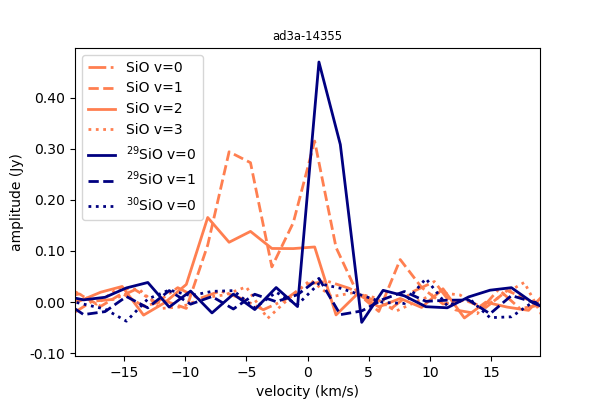}
    \includegraphics[width=0.33\textwidth]{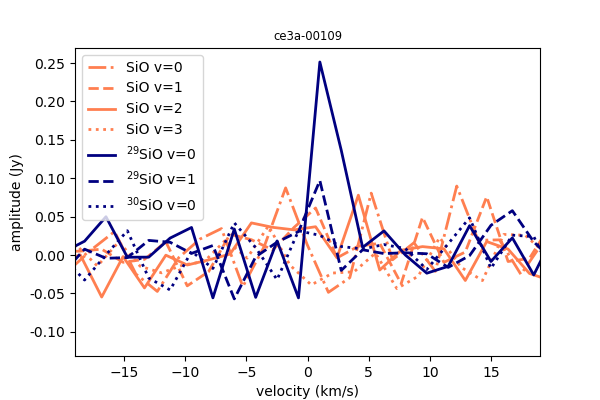}
    \includegraphics[width=0.33\textwidth]{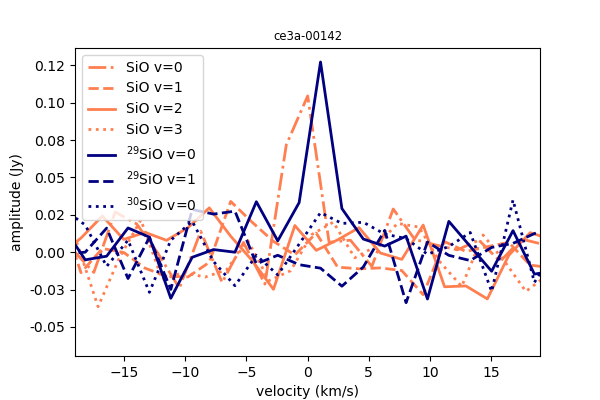}
    \includegraphics[width=0.33\textwidth]{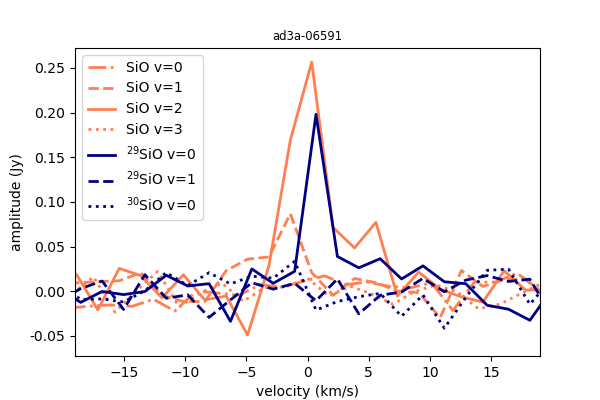}
    \includegraphics[width=0.33\textwidth]{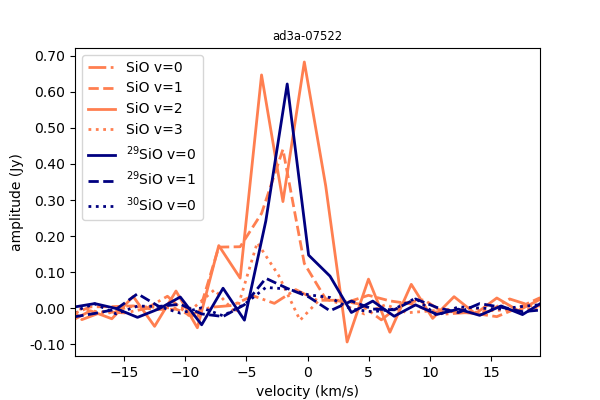}
    \includegraphics[width=0.33\textwidth]{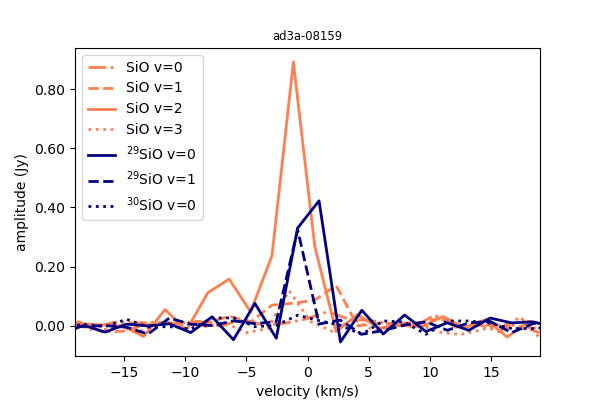}
    \includegraphics[width=0.33\textwidth]{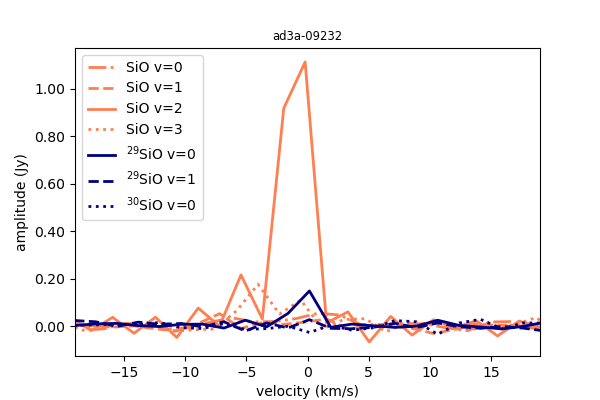}
    \includegraphics[width=0.33\textwidth]{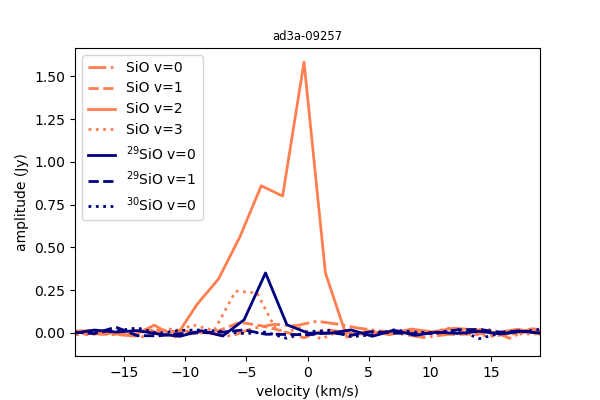}
    \includegraphics[width=0.33\textwidth]{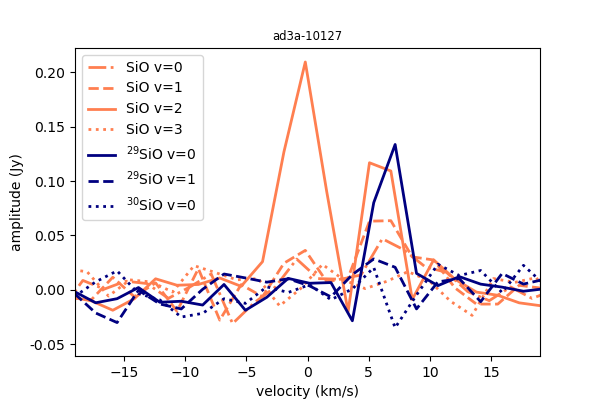}
    \includegraphics[width=0.33\textwidth]{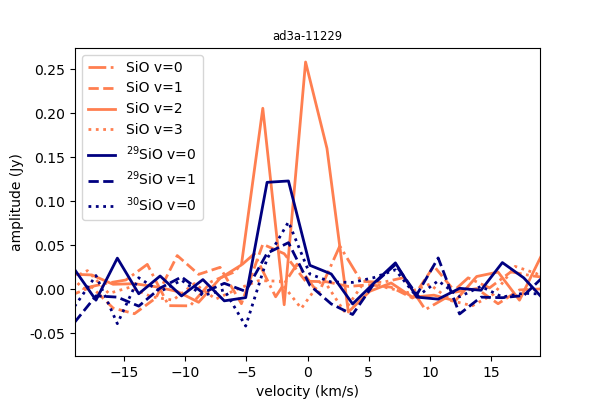}
    \includegraphics[width=0.33\textwidth]{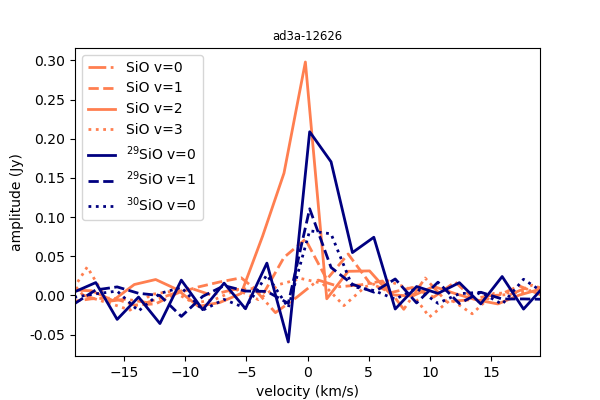}
    \includegraphics[width=0.33\textwidth]{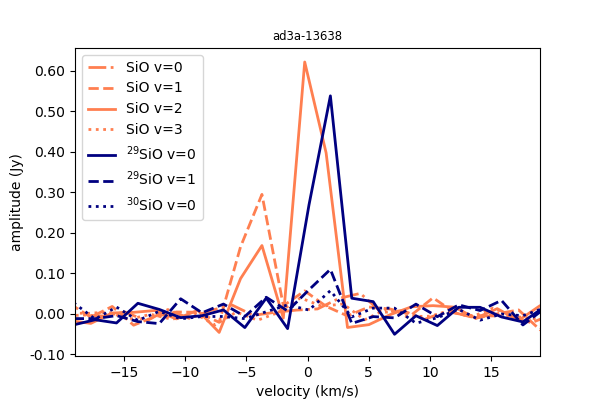}
    \includegraphics[width=0.33\textwidth]{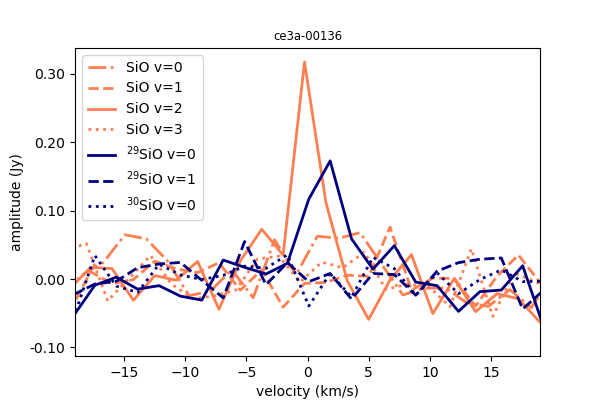}
    \includegraphics[width=0.33\textwidth]{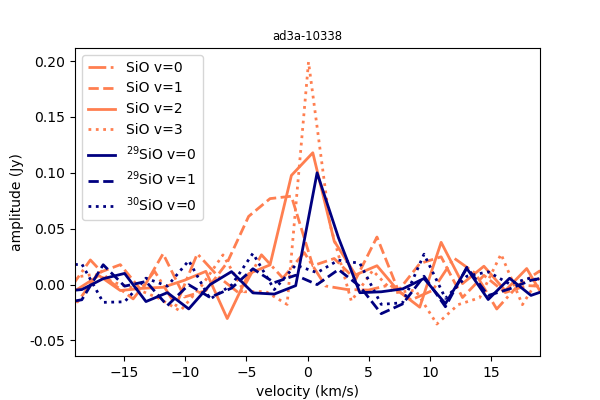}
    \includegraphics[width=0.33\textwidth]{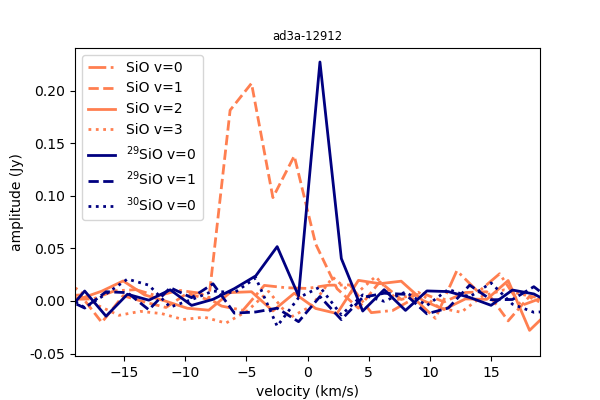}
    \includegraphics[width=0.33\textwidth]{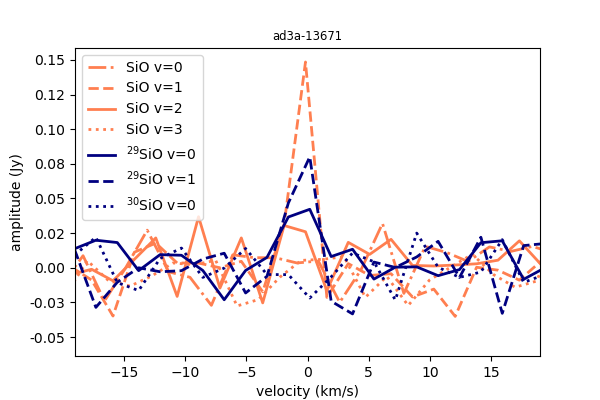}
    $\phantom{\includegraphics[width=0.33\textwidth]{profileplots/lineprofiles_isodom_ad3a-13671.png}}$
\end{figure*}


\section{Full spectra}
\begin{figure*}
\caption{Spectra of the isotopologue-dominated sources. }
    \centering
    \includegraphics[angle=90,height=0.5\textheight]{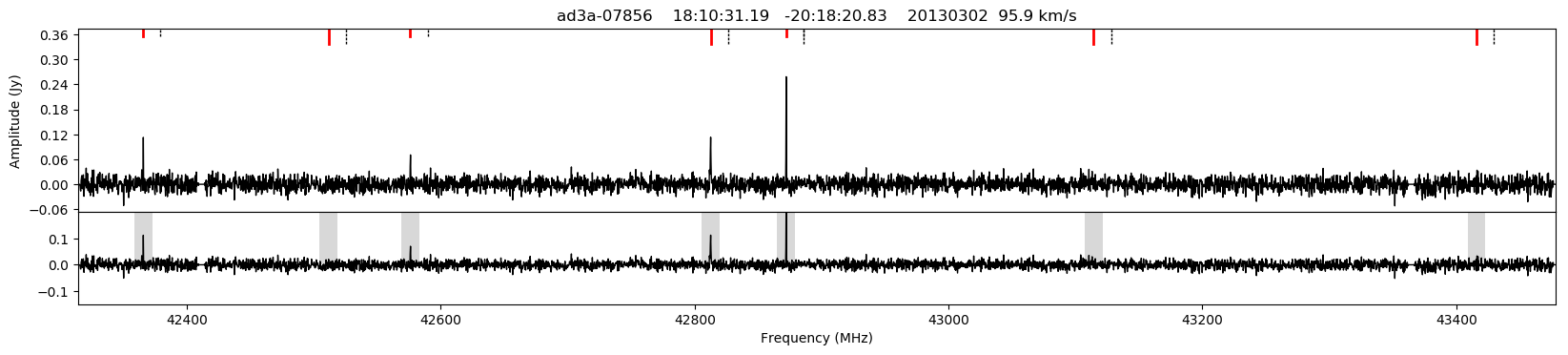}
    \includegraphics[angle=90,height=0.5\textheight]{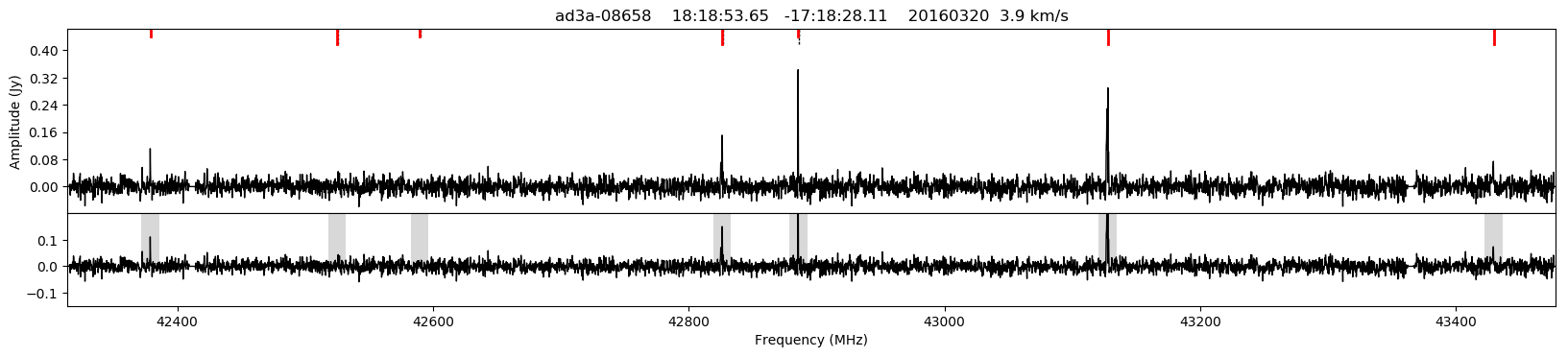}
    \includegraphics[angle=90,height=0.5\textheight]{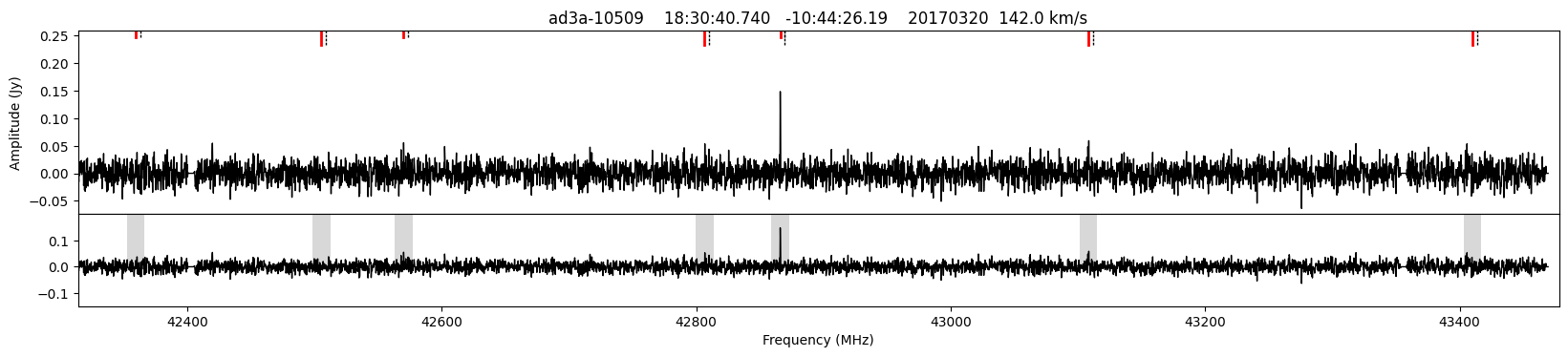}
    \includegraphics[angle=90,height=0.5\textheight]{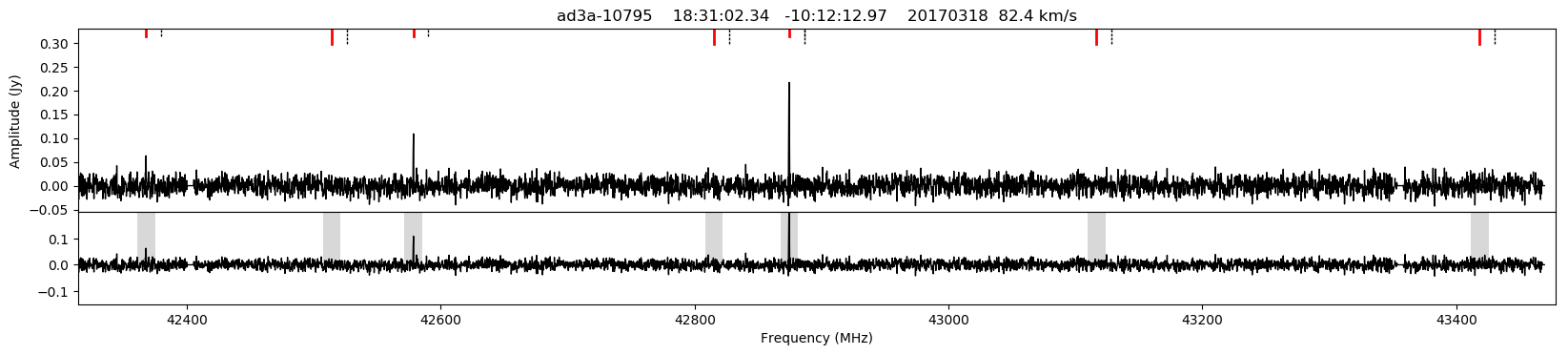}
    \includegraphics[angle=90,height=0.5\textheight]{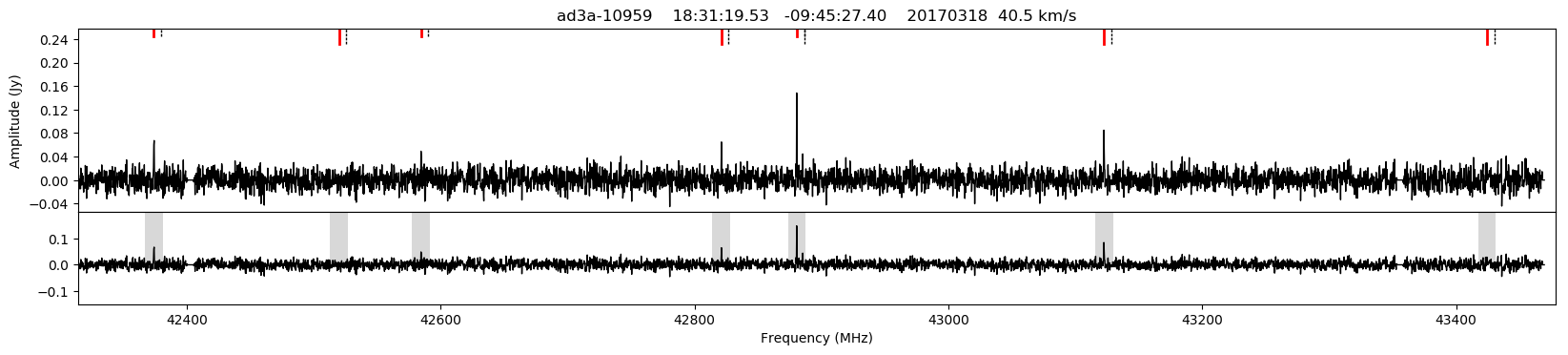}
    \includegraphics[angle=90,height=0.5\textheight]{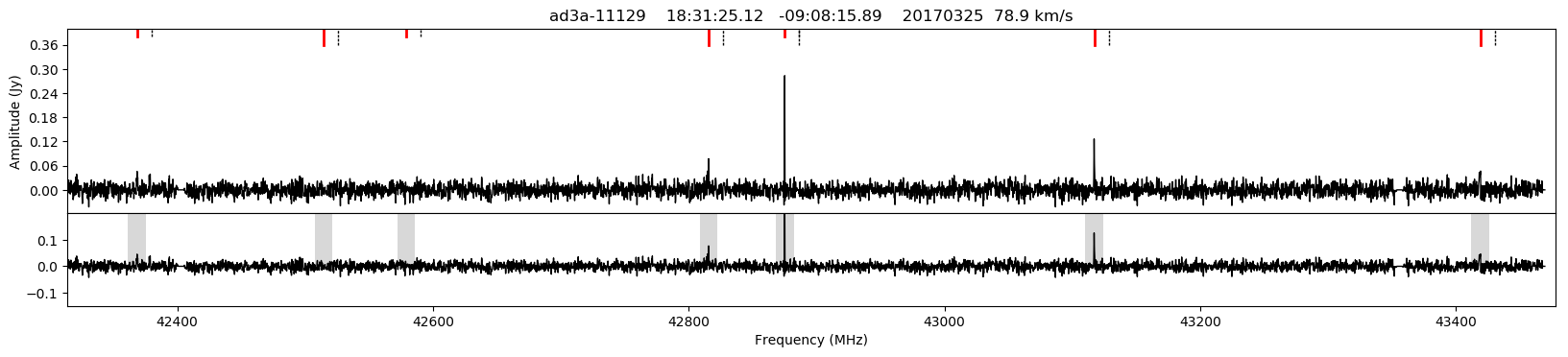}
    \includegraphics[angle=90,height=0.5\textheight]{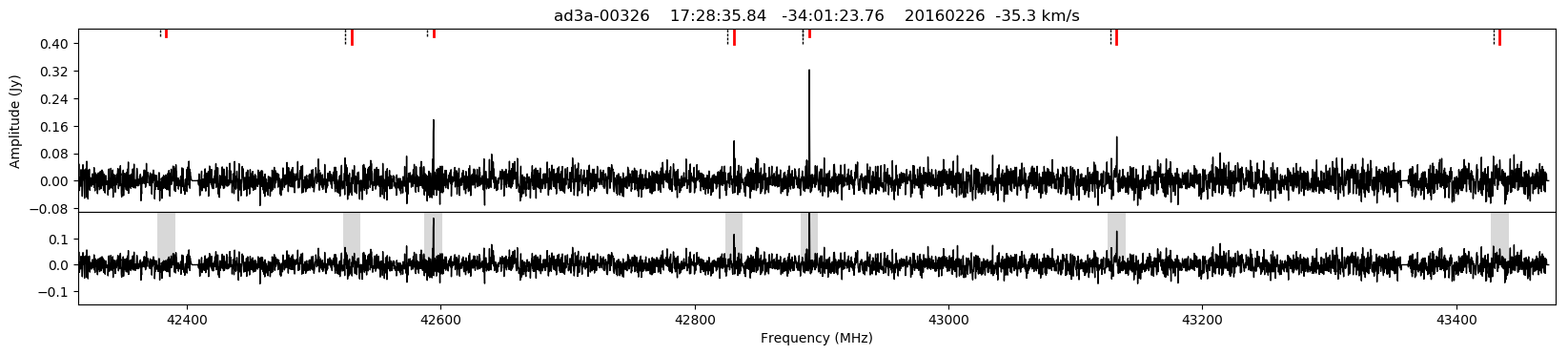}
    \includegraphics[angle=90,height=0.5\textheight]{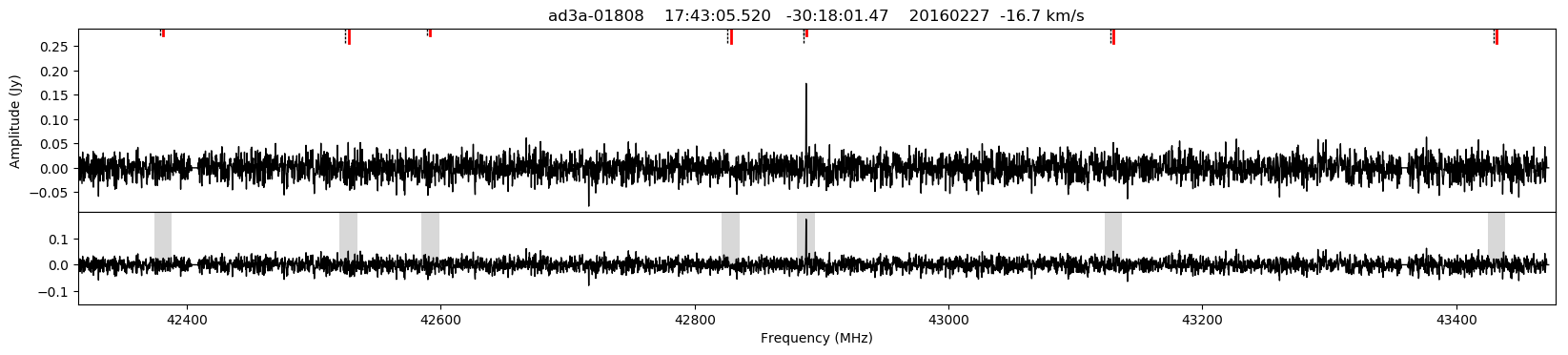}
    \includegraphics[angle=90,height=0.5\textheight]{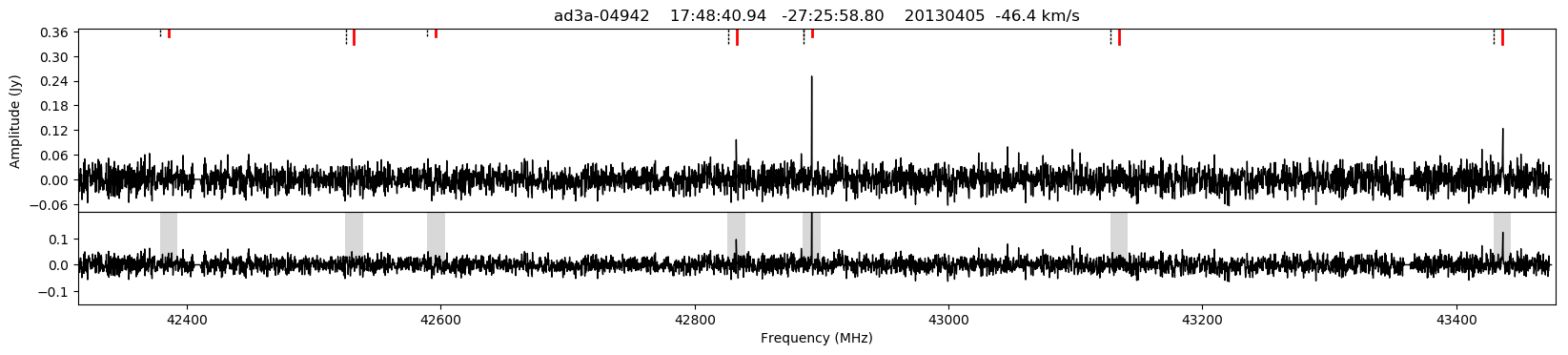}
    \includegraphics[angle=90,height=0.5\textheight]{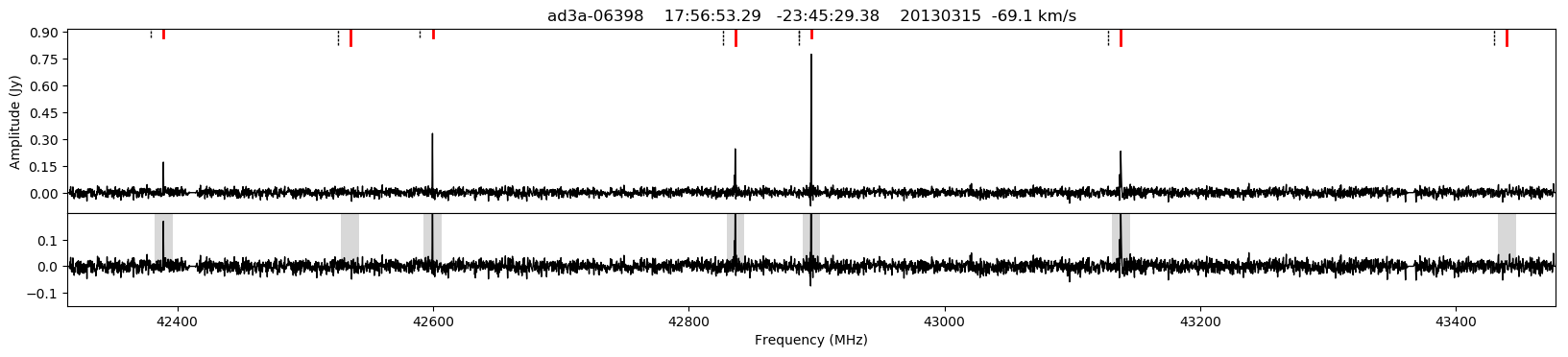}
    \includegraphics[angle=90,height=0.5\textheight]{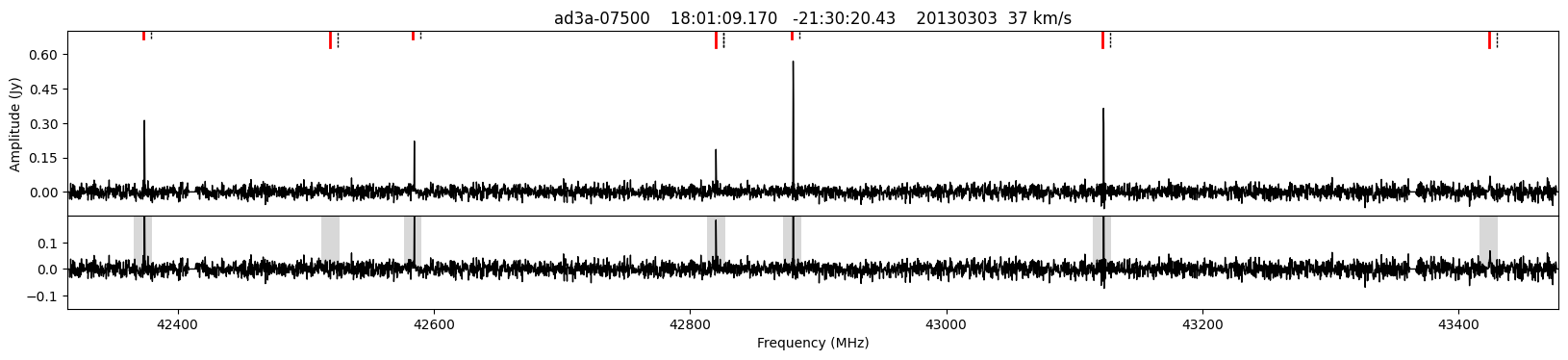}
    \includegraphics[angle=90,height=0.5\textheight]{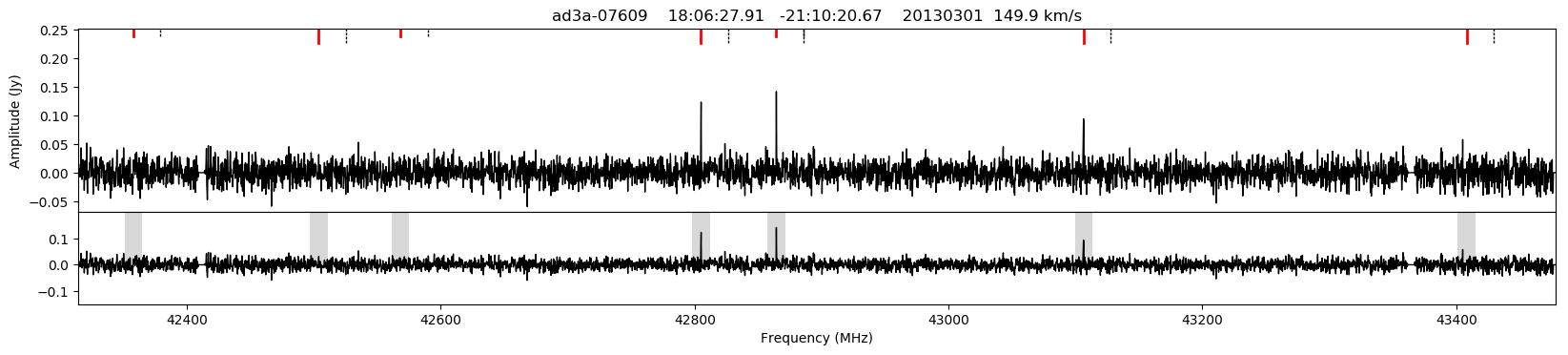}
   \label{fig:allspec2}
\end{figure*}

Figure \ref{fig:allspec2} presents the full, observed spectra of the 35 isotopologue-dominated sources from the BAaDE catalog ordered by Type and BAaDE name (as in Table \ref{tab:baadeoverview}). Name, coordinates, date, and line-of-sight velocity are listed above each spectrum. The top panel shows an auto-scaled spectrum and the bottom shows the same data with a fixed y-axis to emphasize low-signal lines. Grey tick marks in the top panel mark the rest frequencies of the maser transitions listed in Table \ref{tab:lines}, while red tick marks and grey shaded regions show the frequencies of the same transitions shifted by the line-of-sight velocity.

\newpage

\addtocounter{figure}{-1}
\begin{figure*}
    \centering
    \includegraphics[angle=90,height=0.5\textheight]{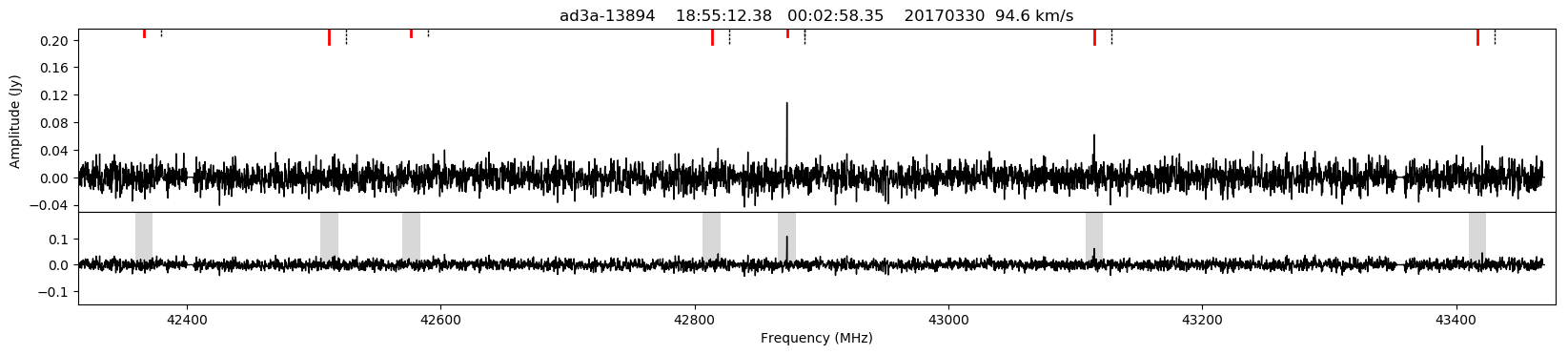}
    \includegraphics[angle=90,height=0.5\textheight]{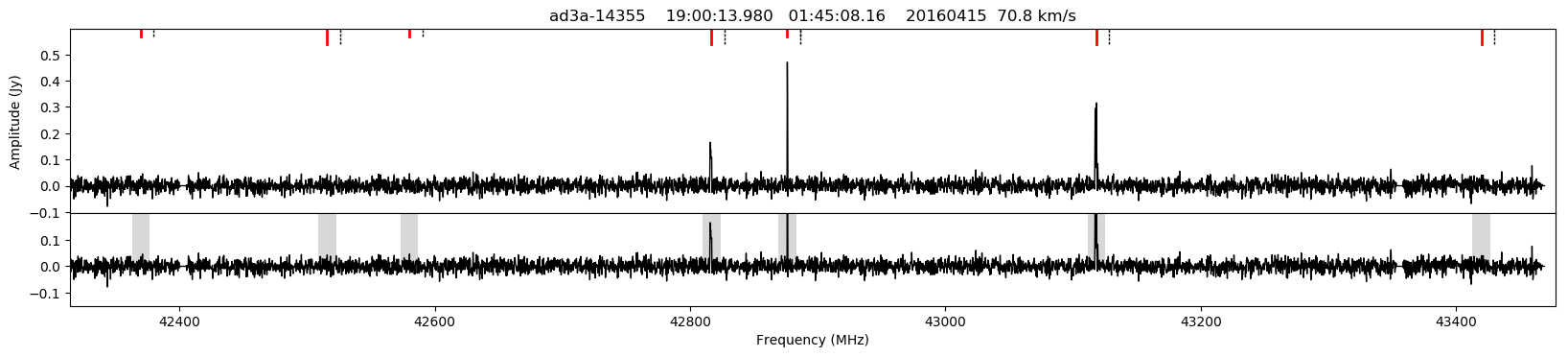}
    \includegraphics[angle=90,height=0.5\textheight]{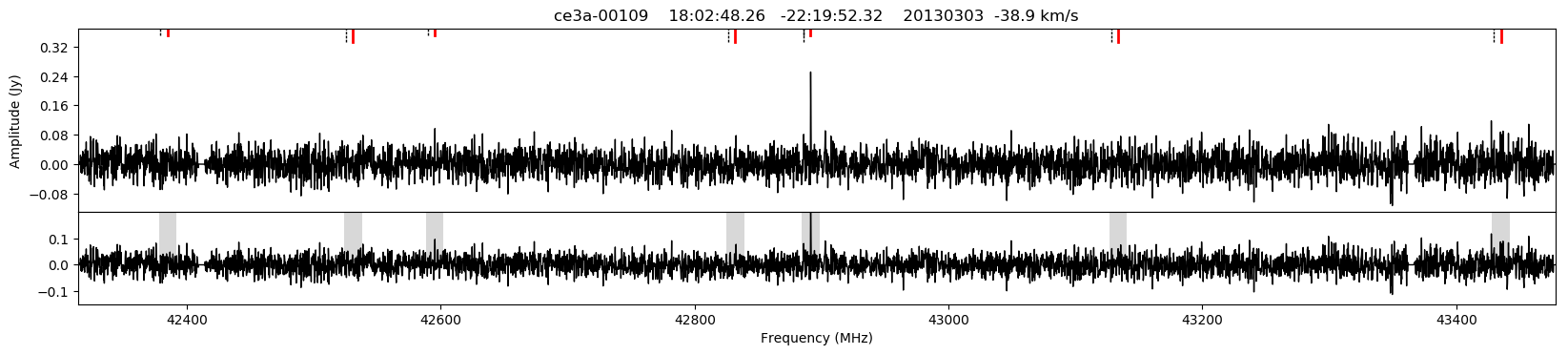}
    \includegraphics[angle=90,height=0.5\textheight]{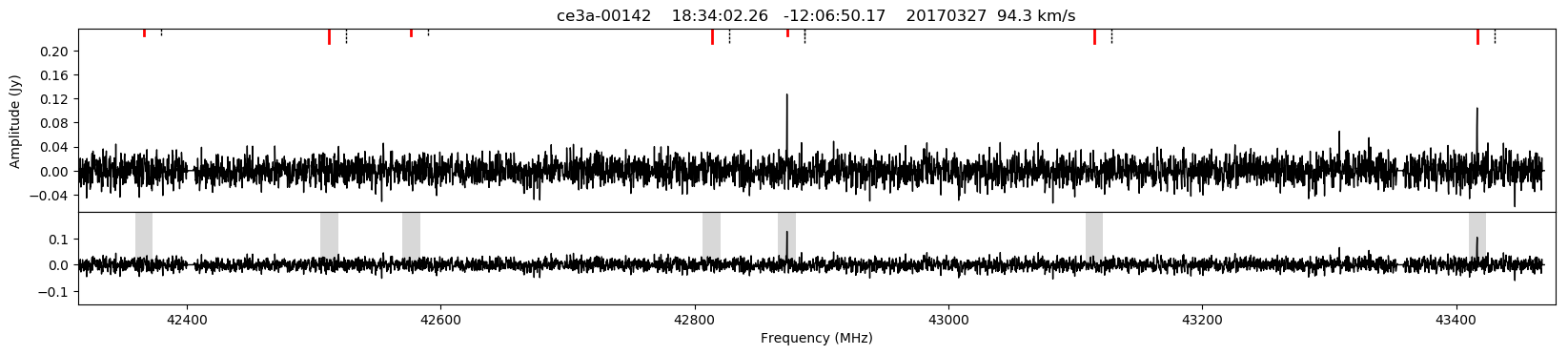}
    \includegraphics[angle=90,height=0.5\textheight]{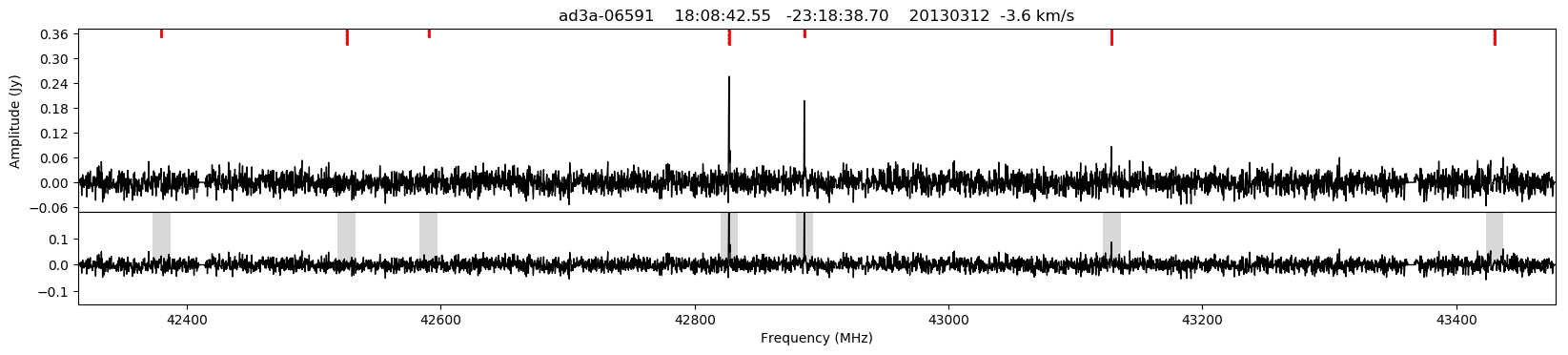}
    \includegraphics[angle=90,height=0.5\textheight]{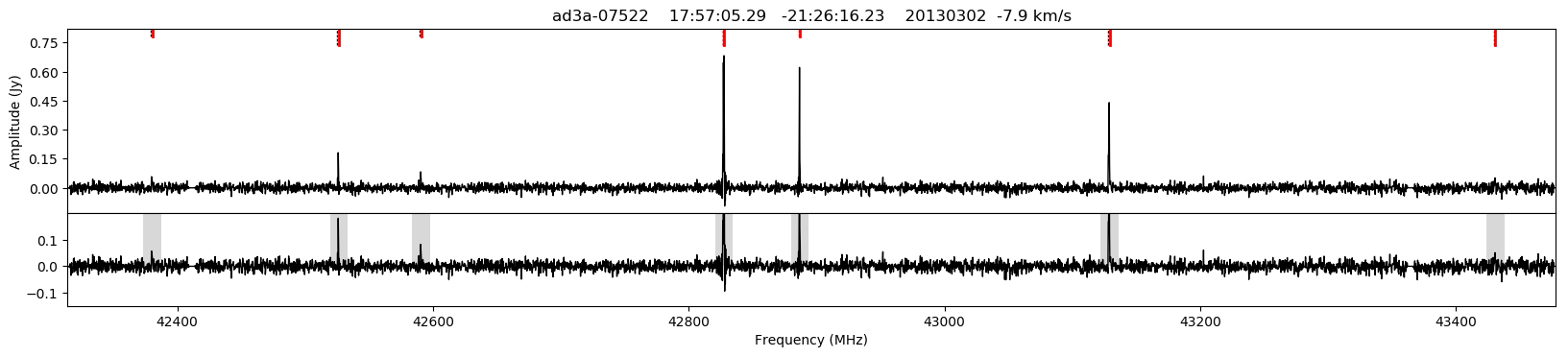}
    \includegraphics[angle=90,height=0.5\textheight]{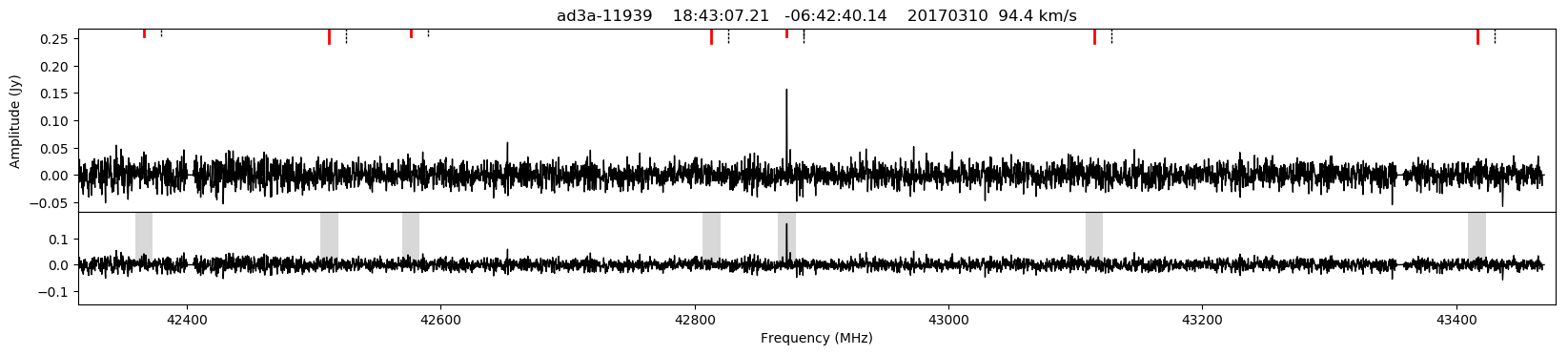}
    \includegraphics[angle=90,height=0.5\textheight]{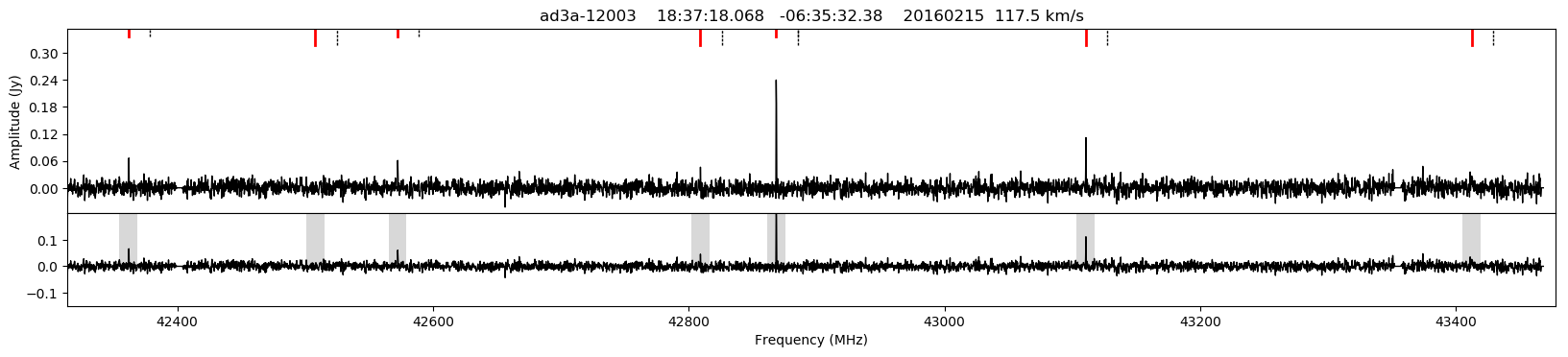}
    \includegraphics[angle=90,height=0.5\textheight]{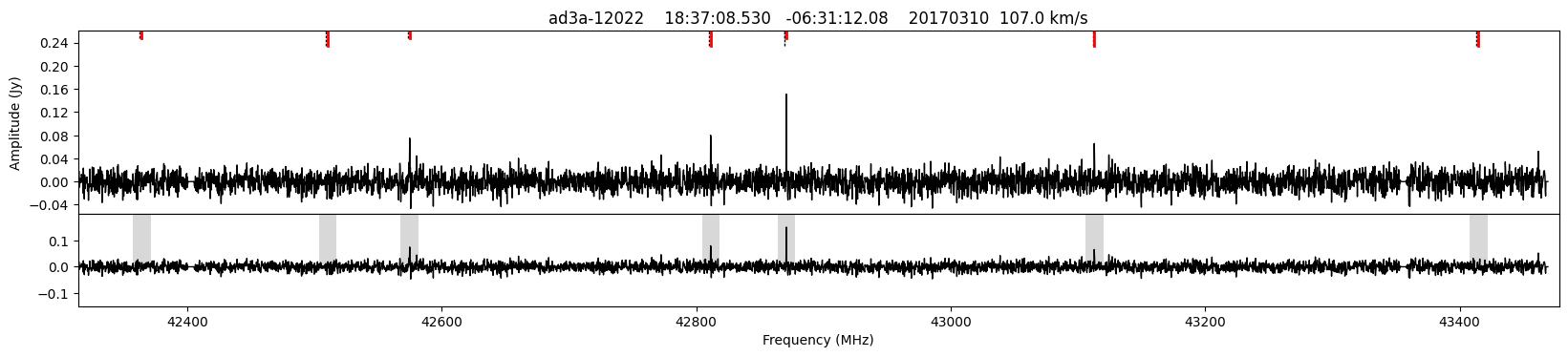}
    \includegraphics[angle=90,height=0.5\textheight]{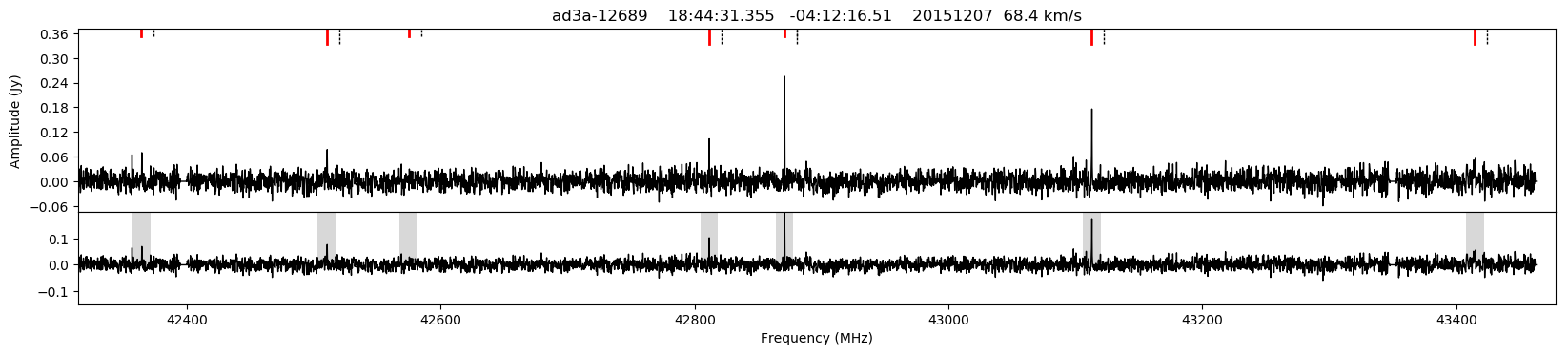}
    \includegraphics[angle=90,height=0.5\textheight]{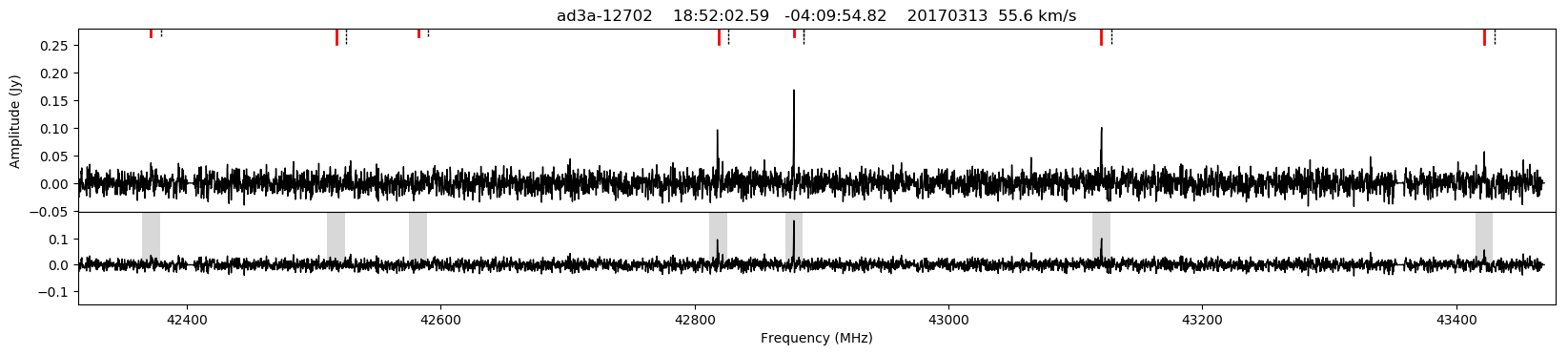}
    \includegraphics[angle=90,height=0.5\textheight]{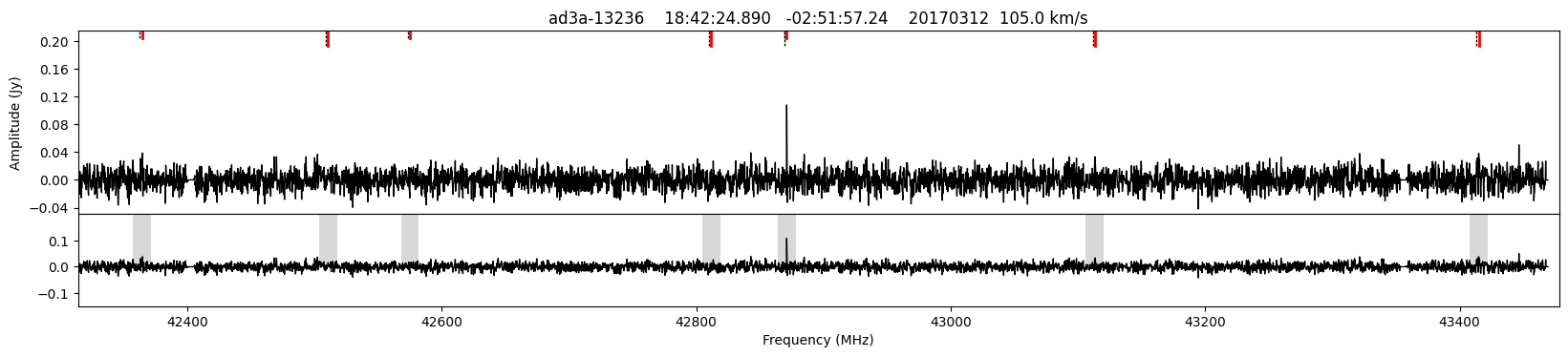}
\end{figure*}
\addtocounter{figure}{-1}
\begin{figure*}
    \centering
    \includegraphics[angle=90,height=0.5\textheight]{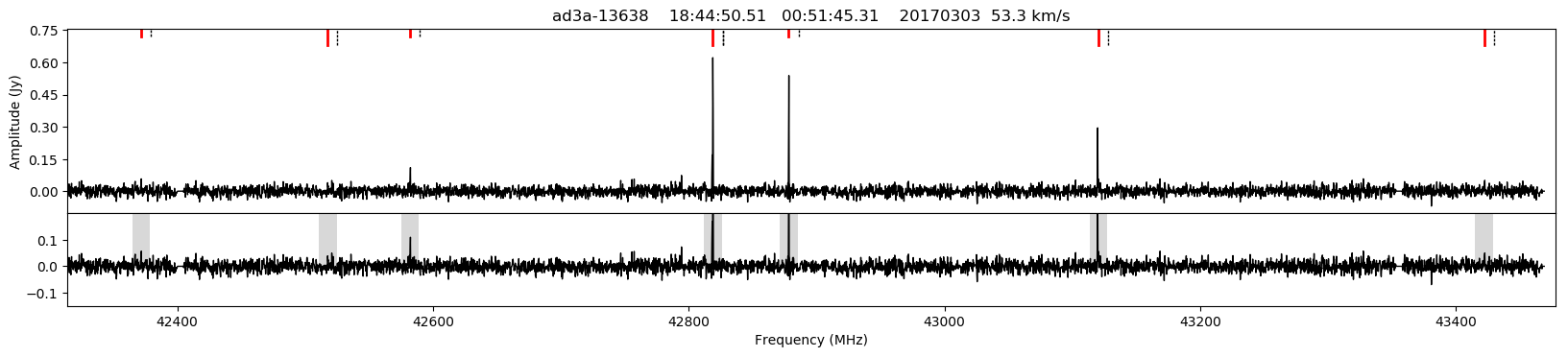}
    \includegraphics[angle=90,height=0.5\textheight]{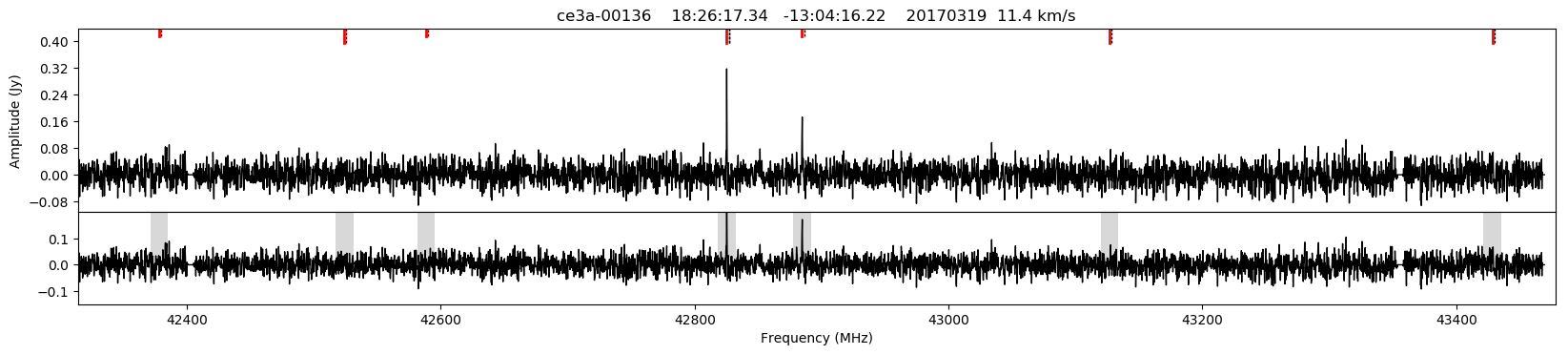}
    \includegraphics[angle=90,height=0.5\textheight]{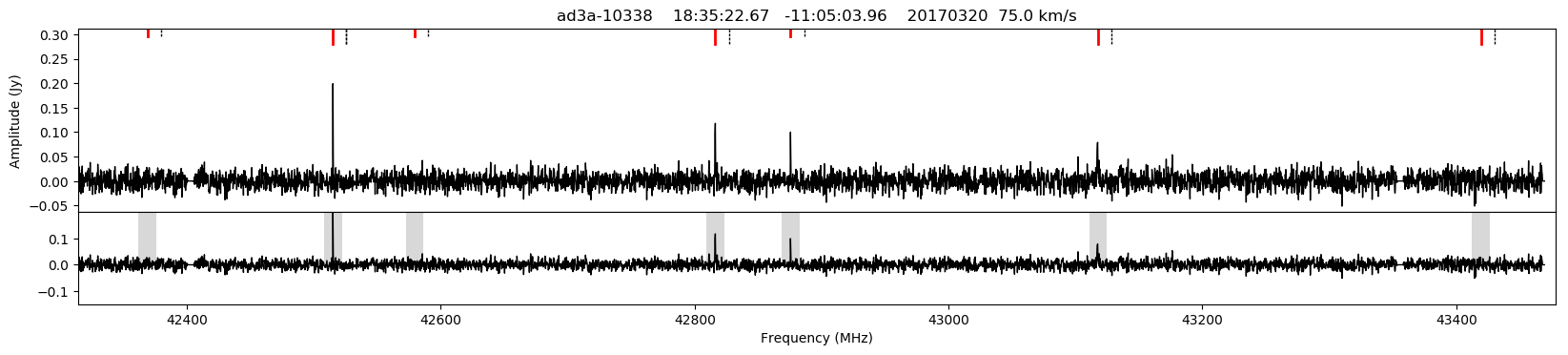}
    \includegraphics[angle=90,height=0.5\textheight]{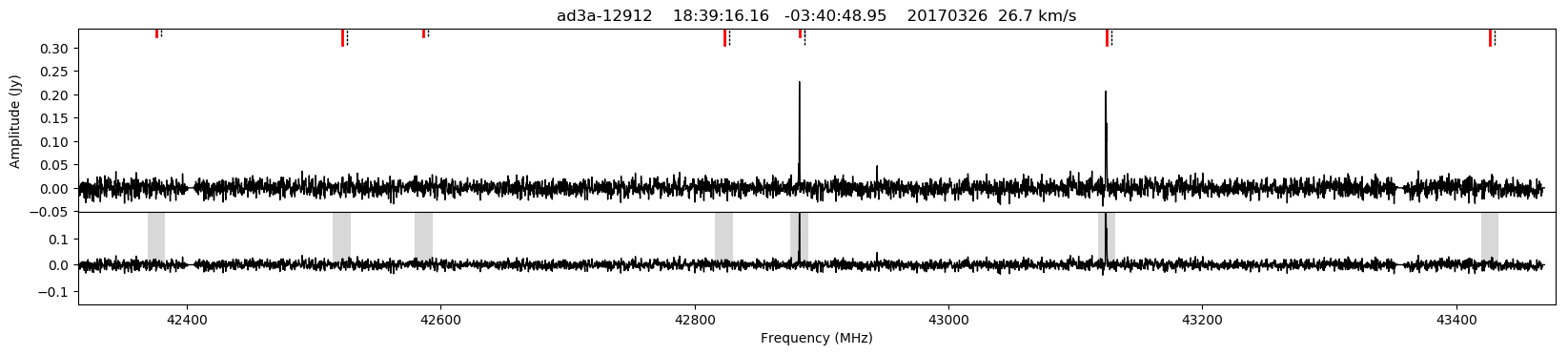}
    \includegraphics[angle=90,height=0.5\textheight]{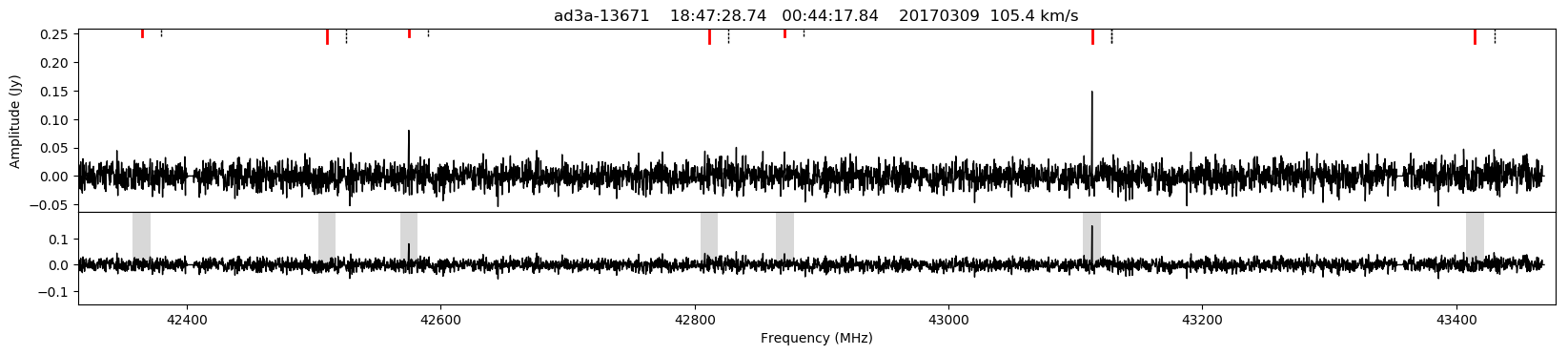}
    $\phantom{\includegraphics[angle=90,height=0.5\textheight]{specs/ad3a-13671.png}}$
    \includegraphics[angle=90,height=0.5\textheight]{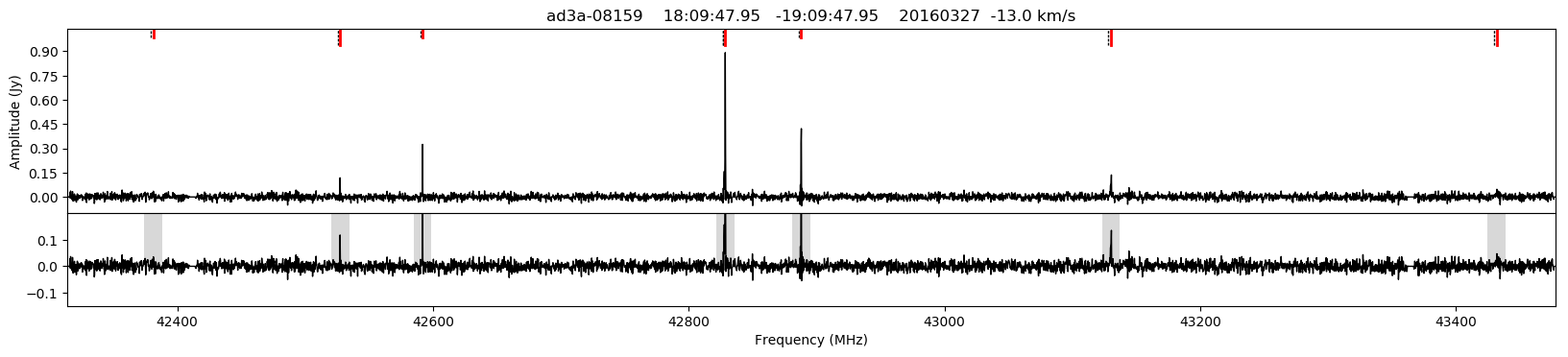}
    \includegraphics[angle=90,height=0.5\textheight]{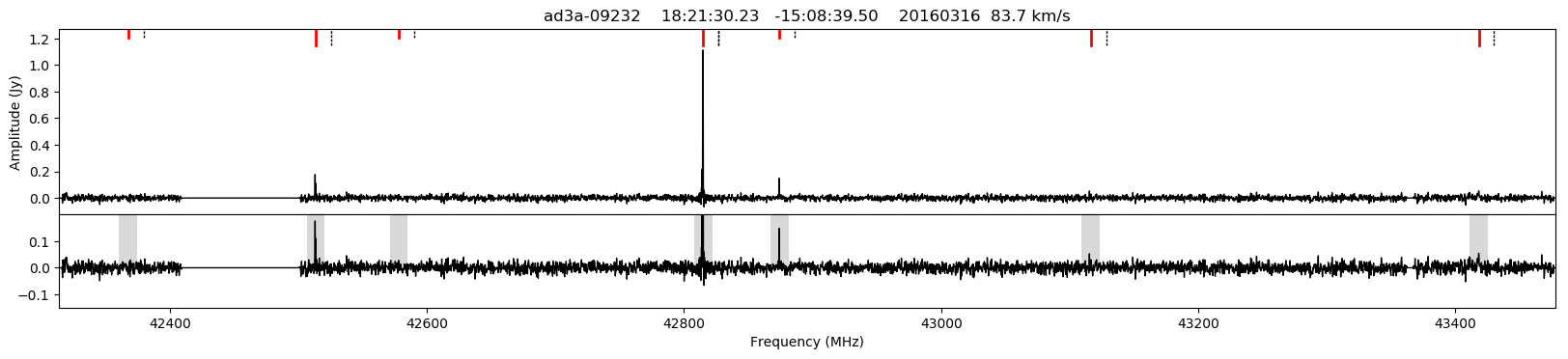}
    \includegraphics[angle=90,height=0.5\textheight]{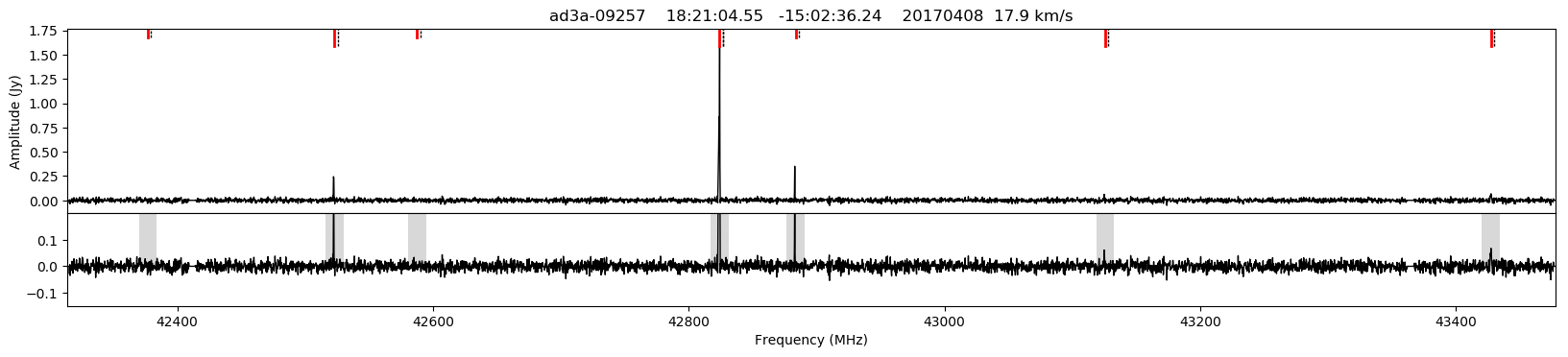}
    \includegraphics[angle=90,height=0.5\textheight]{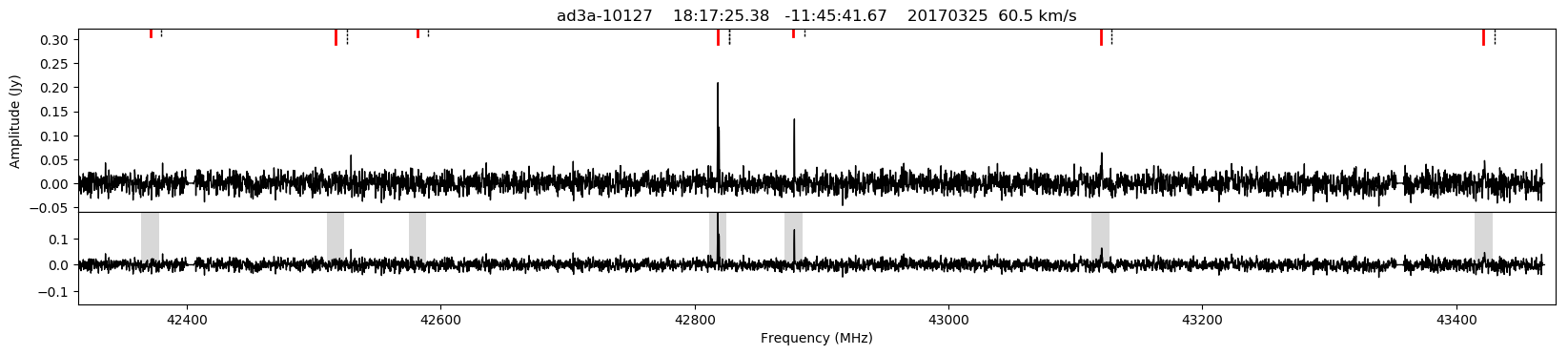}
    \includegraphics[angle=90,height=0.5\textheight]{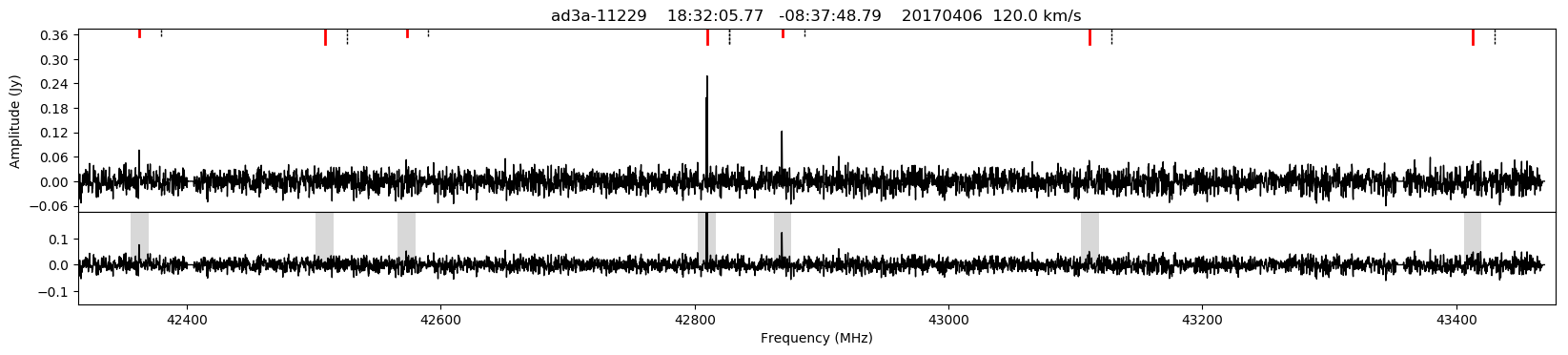}
    \includegraphics[angle=90,height=0.5\textheight]{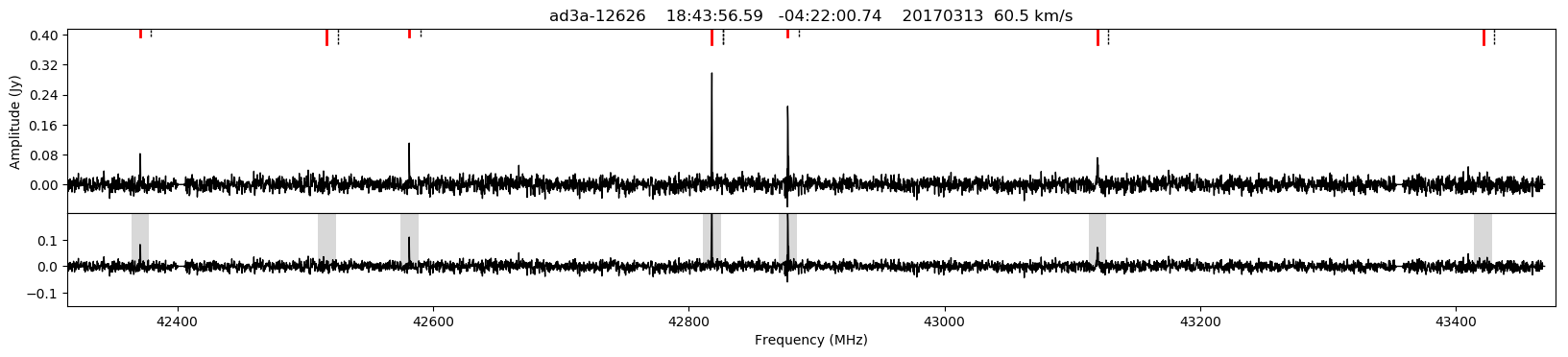}
\end{figure*}

\end{appendix}

\end{document}